\documentclass[11pt]{article}
\usepackage{theapa}

\usepackage[ascii]{inputenc}  %

\usepackage{multirow}
\usepackage{eepic,epic}
\usepackage{epsfig}
\usepackage{graphicx}
\usepackage[table]{xcolor}
\usepackage{longtable,booktabs}

\usepackage{url}

\usepackage{amssymb}
\usepackage{amsfonts}
\usepackage{pifont}
\usepackage{amsmath}
\usepackage{amsthm}

\makeatletter
\renewenvironment{proof}[1][\proofname]{%
  \par\pushQED{\qed}\normalfont%
  \topsep6\p@\@plus6\p@\relax
  \trivlist\item[\hskip\labelsep\bfseries#1\@addpunct{.}]%
  \ignorespaces
}{%
  \popQED\endtrivlist\@endpefalse
}
\makeatother

\usepackage{pifont}
\usepackage{nicefrac}

\newcommand{\condition}{\,\mid \:}

\def\qed{{\ \nolinebreak\hfill\mbox{\qedblob\quad}}}
\newcommand\qedblob{\mbox{\ding{113}}}

\newtheorem{theorem}{Theorem}[section]

\newtheorem{corollary}[theorem]{Corollary}
\newtheorem{definition}[theorem]{Definition}

\newtheorem{corollarytotheproof}[theorem]{Corollary (to the Proof)}

\newcommand{\p}{\ensuremath{\mathrm{P}}}
\newcommand{\np}{\ensuremath{\mathrm{NP}}}
\newcommand{\conp}{\ensuremath{\mathrm{coNP}}}
\newcommand{\coconp}{\ensuremath{\mathrm{cocoNP}}}

\newcommand{\pneqnp}{\ensuremath{\p  \neq \np}}

\newcommand{\itilde}{\ensuremath{\tilde{I}}}

\newcommand{\cale}{\ensuremath{\mathcal E}}

\newcommand{\calt}{\ensuremath{\mathcal T}}
\newcommand{\caltprime}{\ensuremath{\mathcal T}'}

\newcommand{\caledt}{\ensuremath{\cale\hbox{-}\calt}}

\newcommand{\cc}{\ensuremath{\mathrm{CC}}}
\newcommand{\dc}{\ensuremath{\mathrm{DC}}}
\newcommand{\uw}{\ensuremath{\mathrm{UW}}}
\newcommand{\nuw}{\ensuremath{\mathrm{NUW}}}
\newcommand{\pv}{\ensuremath{\mathrm{PV}}}
\newcommand{\pc}{\ensuremath{\mathrm{PC}}}
\newcommand{\rpc}{\ensuremath{\mathrm{RPC}}}
\newcommand{\te}{\ensuremath{\mathrm{TE}}}
\newcommand{\tp}{\ensuremath{\mathrm{TP}}}
\newcommand{\ac}{\ensuremath{\mathrm{AC}}}
\newcommand{\dv}{\ensuremath{\mathrm{DV}}}
\newcommand{\av}{\ensuremath{\mathrm{AV}}}
\newcommand{\uac}{\ensuremath{\mathrm{UAC}}}

\newcommand{\ccpvteuw}{\cc\hbox{-}\pv\hbox{-}\allowbreak\te\hbox{-}\allowbreak\uw}
\newcommand{\ccpvtenuw}{\cc\hbox{-}\pv\hbox{-}\allowbreak\te\hbox{-}\allowbreak\nuw}
\newcommand{\ccpvtpuw}{\cc\hbox{-}\pv\hbox{-}\allowbreak\tp\hbox{-}\allowbreak\uw}
\newcommand{\ccpvtpnuw}{\cc\hbox{-}\pv\hbox{-}\allowbreak\tp\hbox{-}\allowbreak\nuw}
\newcommand{\ccpcteuw}{\cc\hbox{-}\pc\hbox{-}\allowbreak\te\hbox{-}\allowbreak\uw}
\newcommand{\ccpctenuw}{\cc\hbox{-}\pc\hbox{-}\allowbreak\te\hbox{-}\allowbreak\nuw}
\newcommand{\ccpctpuw}{\cc\hbox{-}\pc\hbox{-}\allowbreak\tp\hbox{-}\allowbreak\uw}
\newcommand{\ccpctpnuw}{\cc\hbox{-}\pc\hbox{-}\allowbreak\tp\hbox{-}\allowbreak\nuw}
\newcommand{\ccrpcteuw}{\cc\hbox{-}\rpc\hbox{-}\allowbreak\te\hbox{-}\allowbreak\uw}
\newcommand{\ccrpctenuw}{\cc\hbox{-}\rpc\hbox{-}\allowbreak\te\hbox{-}\allowbreak\nuw}
\newcommand{\ccrpctpuw}{\cc\hbox{-}\rpc\hbox{-}\allowbreak\tp\hbox{-}\allowbreak\uw}
\newcommand{\ccrpctpnuw}{\cc\hbox{-}\rpc\hbox{-}\allowbreak\tp\hbox{-}\allowbreak\nuw}

\newcommand{\dcpvtpuw}{\dc\hbox{-}\pv\hbox{-}\allowbreak\tp\hbox{-}\allowbreak\uw}
\newcommand{\dcpvtpnuw}{\dc\hbox{-}\pv\hbox{-}\allowbreak\tp\hbox{-}\allowbreak\nuw}
\newcommand{\dcpcteuw}{\dc\hbox{-}\pc\hbox{-}\allowbreak\te\hbox{-}\allowbreak\uw}
\newcommand{\dcpctenuw}{\dc\hbox{-}\pc\hbox{-}\allowbreak\te\hbox{-}\allowbreak\nuw}
\newcommand{\dcpctpuw}{\dc\hbox{-}\pc\hbox{-}\allowbreak\tp\hbox{-}\allowbreak\uw}
\newcommand{\dcpctpnuw}{\dc\hbox{-}\pc\hbox{-}\allowbreak\tp\hbox{-}\allowbreak\nuw}
\newcommand{\dcrpcteuw}{\dc\hbox{-}\rpc\hbox{-}\allowbreak\te\hbox{-}\allowbreak\uw}
\newcommand{\dcrpctenuw}{\dc\hbox{-}\rpc\hbox{-}\allowbreak\te\hbox{-}\allowbreak\nuw}
\newcommand{\dcrpctpuw}{\dc\hbox{-}\rpc\hbox{-}\allowbreak\tp\hbox{-}\allowbreak\uw}
\newcommand{\dcrpctpnuw}{\dc\hbox{-}\rpc\hbox{-}\allowbreak\tp\hbox{-}\allowbreak\nuw}

\newcommand{\ccuacnuw}{\cc\hbox{-}\allowbreak\uac\hbox{-}\allowbreak\nuw}
\newcommand{\ccacnuw}{\cc\hbox{-}\allowbreak\ac\hbox{-}\allowbreak\nuw}
\newcommand{\ccdcnuw}{\cc\hbox{-}\allowbreak\dc\hbox{-}\allowbreak\nuw}
\newcommand{\ccdvnuw}{\cc\hbox{-}\allowbreak\dv\hbox{-}\allowbreak\nuw}
\newcommand{\ccavnuw}{\cc\hbox{-}\allowbreak\av\hbox{-}\allowbreak\nuw}

\newcommand{\dcdcnuw}{\dc\hbox{-}\allowbreak\dc\hbox{-}\allowbreak\nuw}
\newcommand{\dcdvnuw}{\dc\hbox{-}\allowbreak\dv\hbox{-}\allowbreak\nuw}

\newcommand{\dcdcuw}{\dc\hbox{-}\allowbreak\dc\hbox{-}\allowbreak\uw}
\newcommand{\dcdvuw}{\dc\hbox{-}\allowbreak\dv\hbox{-}\allowbreak\uw}

\newcommand{\caledash}{\ensuremath{\cale\hbox{-}}}
\newcommand{\pluralitydash}{\mathrm{Plurality}\hbox{-}}
\newcommand{\vetodash}{\mathrm{Veto}\hbox{-}}
\newcommand{\approvaldash}{\mathrm{Approval}\hbox{-}}

\newcommand{\pluralitydcpcteuw}{\pluralitydash\dc\hbox{-}\allowbreak\pc\hbox{-}\allowbreak\te\hbox{-}\allowbreak\uw}
\newcommand{\pluralitydcpctenuw}{\pluralitydash\dc\hbox{-}\allowbreak\pc\hbox{-}\allowbreak\te\hbox{-}\allowbreak\nuw}

\newcommand{\pluralitydcpctpnuw}{\pluralitydash\dc\hbox{-}\allowbreak\pc\hbox{-}\allowbreak\tp\hbox{-}\allowbreak\nuw}
\newcommand{\pluralitydcrpcteuw}{\pluralitydash\dc\hbox{-}\allowbreak\rpc\hbox{-}\allowbreak\te\hbox{-}\allowbreak\uw}

\newcommand{\vetodcpvteuw}{\vetodash\dc\hbox{-}\allowbreak\pv\hbox{-}\allowbreak\te\hbox{-}\allowbreak\uw}
\newcommand{\vetodcpvtenuw}{\vetodash\dc\hbox{-}\allowbreak\pv\hbox{-}\allowbreak\te\hbox{-}\allowbreak\nuw}

\newcommand{\vetodcpcteuw}{\vetodash\dc\hbox{-}\allowbreak\pc\hbox{-}\allowbreak\te\hbox{-}\allowbreak\uw}
\newcommand{\vetodcpctenuw}{\vetodash\dc\hbox{-}\allowbreak\pc\hbox{-}\allowbreak\te\hbox{-}\allowbreak\nuw}

\newcommand{\vetodcpctpnuw}{\vetodash\dc\hbox{-}\allowbreak\pc\hbox{-}\allowbreak\tp\hbox{-}\allowbreak\nuw}
\newcommand{\vetodcrpcteuw}{\vetodash\dc\hbox{-}\allowbreak\rpc\hbox{-}\allowbreak\te\hbox{-}\allowbreak\uw}

\newcommand{\approvalccpcteuw}{\approvaldash\cc\hbox{-}\allowbreak\pc\hbox{-}\allowbreak\te\hbox{-}\allowbreak\uw}
\newcommand{\approvalccpctenuw}{\approvaldash\cc\hbox{-}\allowbreak\pc\hbox{-}\allowbreak\te\hbox{-}\allowbreak\nuw}
\newcommand{\approvalccpctpuw}{\approvaldash\cc\hbox{-}\allowbreak\pc\hbox{-}\allowbreak\tp\hbox{-}\allowbreak\uw}
\newcommand{\approvalccpctpnuw}{\approvaldash\cc\hbox{-}\allowbreak\pc\hbox{-}\allowbreak\tp\hbox{-}\allowbreak\nuw}
\newcommand{\approvalccrpcteuw}{\approvaldash\cc\hbox{-}\allowbreak\rpc\hbox{-}\allowbreak\te\hbox{-}\allowbreak\uw}
\newcommand{\approvalccrpctenuw}{\approvaldash\cc\hbox{-}\allowbreak\rpc\hbox{-}\allowbreak\te\hbox{-}\allowbreak\nuw}
\newcommand{\approvalccrpctpuw}{\approvaldash\cc\hbox{-}\allowbreak\rpc\hbox{-}\allowbreak\tp\hbox{-}\allowbreak\uw}
\newcommand{\approvalccrpctpnuw}{\approvaldash\cc\hbox{-}\allowbreak\rpc\hbox{-}\allowbreak\tp\hbox{-}\allowbreak\nuw}
\newcommand{\approvaldcpvteuw}{\approvaldash\dc\hbox{-}\allowbreak\pv\hbox{-}\allowbreak\te\hbox{-}\allowbreak\uw}
\newcommand{\approvaldcpvtenuw}{\approvaldash\dc\hbox{-}\allowbreak\pv\hbox{-}\allowbreak\te\hbox{-}\allowbreak\nuw}
\newcommand{\approvaldcpvtpuw}{\approvaldash\dc\hbox{-}\allowbreak\pv\hbox{-}\allowbreak\tp\hbox{-}\allowbreak\uw}

\newcommand{\approvaldcpcteuw}{\approvaldash\dc\hbox{-}\allowbreak\pc\hbox{-}\allowbreak\te\hbox{-}\allowbreak\uw}
\newcommand{\approvaldcpctenuw}{\approvaldash\dc\hbox{-}\allowbreak\pc\hbox{-}\allowbreak\te\hbox{-}\allowbreak\nuw}
\newcommand{\approvaldcpctpuw}{\approvaldash\dc\hbox{-}\allowbreak\pc\hbox{-}\allowbreak\tp\hbox{-}\allowbreak\uw}
\newcommand{\approvaldcpctpnuw}{\approvaldash\dc\hbox{-}\allowbreak\pc\hbox{-}\allowbreak\tp\hbox{-}\allowbreak\nuw}
\newcommand{\approvaldcrpcteuw}{\approvaldash\dc\hbox{-}\allowbreak\rpc\hbox{-}\allowbreak\te\hbox{-}\allowbreak\uw}
\newcommand{\approvaldcrpctenuw}{\approvaldash\dc\hbox{-}\allowbreak\rpc\hbox{-}\allowbreak\te\hbox{-}\allowbreak\nuw}
\newcommand{\approvaldcrpctpuw}{\approvaldash\dc\hbox{-}\allowbreak\rpc\hbox{-}\allowbreak\tp\hbox{-}\allowbreak\uw}
\newcommand{\approvaldcrpctpnuw}{\approvaldash\dc\hbox{-}\allowbreak\rpc\hbox{-}\allowbreak\tp\hbox{-}\allowbreak\nuw}

\newcommand{\approvaldcdcnuw}{\approvaldash\dc\hbox{-}\allowbreak\dc\hbox{-}\allowbreak\nuw}
\newcommand{\approvaldcdvnuw}{\approvaldash\dc\hbox{-}\allowbreak\dv\hbox{-}\allowbreak\nuw}

\newcommand{\approvaldcdcuw}{\approvaldash\dc\hbox{-}\allowbreak\dc\hbox{-}\allowbreak\uw}
\newcommand{\approvaldcdvuw}{\approvaldash\dc\hbox{-}\allowbreak\dv\hbox{-}\allowbreak\uw}

\usepackage[position=top]{caption}
\newcount\mylineno
\makeatletter
\newcommand{\LTautolinebreak}[1][\@nil]{\\}
\makeatother

\usepackage{array} 
\newcolumntype{P}[1]{>{\centering\arraybackslash}p{#1}}

\makeatletter
\newcommand\smalltblfont{\@setfontsize\smalltblfont{8}{9}}
\makeatother

\begin{document}\sloppy

\title{
Separating and Collapsing 
Electoral Control Types\thanks{Supported in part
    by NSF grants CCF-2006496
and DUE-2135431,
    and CIFellows grant CIF2020-UR-36.}}

\author{%
    Benjamin Carleton\thanks{Work done in part while at the University of Rochester's Department of Computer Science.}\\
    Department of Computer Science\\Cornell University\\Ithaca, NY 14850
    \and
    Michael C. Chavrimootoo and  
    Lane A. Hemaspaandra\\
    Department of Computer Science\\University of Rochester\\Rochester, NY 14627
    \and
    David E. Narv\'{a}ez\footnotemark[2]\\
    Bradley Department of Electrical and Computer Engineering\\
    Virginia Tech\\
    Blacksburg, VA 24061
    \and
    Conor Taliancich\footnotemark[2]\\
    Property Matrix\\
    Culver City, CA 90230 
    \and
    Henry B. Welles\\
    Department of Computer Science\\University of Rochester\\Rochester, NY 14627
}

\date{July 2, 2022; revised May 30, 2024}
\maketitle

\begin{abstract}
Electoral control refers to attacking elections by
adding, deleting, or partitioning voters or candidates~\cite{bar-tov-tri:j:control}.
It was recently discovered, for seven pairs $(\calt,\calt')$ of seemingly
  distinct standard electoral control types, that $\calt$ and $\calt'$ are in practice
  identical: 
  For each input $I$ and each election system $\cale$, 
  $I$ is a ``yes'' instance of both $\calt$ and $\calt'$ under $\cale$, or of
  neither~\cite{hem-hem-men:j:search-versus-decision}. 
Surprisingly, this had previously gone undetected 
even as the field was score-carding how many
  standard control types various election systems were resistant to; various
  ``different'' 
  cells 
  on such score cards were, unknowingly, duplicate effort
  on the same issue. 
 This naturally raises the worry that perhaps
 other
  pairs of control types 
  are identical, and so work still is
  being needlessly duplicated.

  We completely determine, for all
  standard control types, which pairs are, for elections
  whose votes are 
  linear orderings of the candidates, always identical.
  In particular, we prove that 
  no identical control pairs exist beyond
  the known seven.
We also 
for three central
  election systems completely determine which control pairs are
  identical (``collapse'') with respect to those particular election systems, and we also explore
  containment and incomparability relationships between control pairs.  
  For approval
  voting, which has a different ``type'' for its votes,  \citeS{hem-hem-men:j:search-versus-decision} %
  seven collapses still hold (since we observe that their argument applies to all
  election systems). 
  However, we find 14 additional collapses
  that hold for approval voting but do not hold for some election systems whose
  votes are 
  linear orderings of the candidates.  
  We 
  find one 
  new
  collapse for veto elections and none
  for plurality.
  We prove that 
  each 
  of the three election systems mentioned have no collapses other than those 
  inherited from %
  \citeA{hem-hem-men:j:search-versus-decision}
  or added in the present paper.
We
  establish many new containment relationships 
  between 
  separating control pairs, and for each separating pair 
  of standard 
  control
  types classify 
  its separation in terms of either 
  containment (always, and strict on some inputs) or incomparability.

Our work, for the general case and 
these 
three 
important election systems,  clarifies the landscape of 
the 44 standard control types, for each pair 
 collapsing or separating them, and also providing finer-grained information 
on 
the separations.
\end{abstract}

\section{Introduction}\label{sec:introduction}

Preference aggregation is a central focus of 
social choice theory, while a central focus of AI is the ways 
in which agents interact, often in tasks or decision-making 
settings. It thus is natural that ever since 
computational social choice theory emerged as a field, 
it has been prominent at AI venues.
The rise of online decision-making and the 
increasing 
use of agents in our 
increasingly online world make %
computational social choice, and in particular 
its approach to the study of preference aggregation---including 
its focus on the study of the complexity of attacks on elections---%
ever more %
important.
This paper explores, and 
for some central cases 
definitively determines, the extent of a type of 
co-incidence %
(%
namely,
the
equality of seemingly different notions)
that affects one of the most important models---control attacks---in the richly studied and important area of the 
(non)resistance %
to attacks of preference aggregation by voting.
This is important in computational social choice and to AI, 
since finding collapses of 
two control types for a given election system 
means the two cases are effectively identical as sets, and so saves 
researchers from doing 
the duplicate work of separately studying 
the control types 
as to
any set-based property, such as complexity, etc.

Let us start with 
a brief 
parallel.
Suppose Professor Fou specializes in complexity classification, and for each 
problem that comes through the door,
Professor Fou tries to prove it complete 
for 
Fou's pet list of classes: $\np$, $\conp$, and $\coconp$. 
So on Monday, a problem ${\rm P}_1$ from scheduling theory would come in the 
door and
Fou might prove it complete for $\conp$ and would give arguments suggesting
$\np$-completeness and $\coconp$-completeness are each unlikely. On Tuesday, a problem 
${\rm P}_2$ from
computational social choice might come in the door and Fou might prove it
complete for $\np$ and $\coconp$, and would give evidence suggesting 
$\conp$-completeness is wildly unlikely.  And so on for many years.

Most CS academics would be horrified at the situation and would say: How 
can
Fou not have stepped back and tried to see the big picture---and 
realized that $\np$ and $\coconp$ are exactly the same class, and thus that Fou was 
doing duplicate work! 

This seems at first a comic situation, yet a more subtle cousin of it has been 
in play in the computational social choice world for many years.

Let us explain what we mean by that. Control and manipulation are the two 
families of attack types that were the focus of the seminal papers of Bartholdi,
Tovey, and Trick~\citeyear{bar-tov-tri:j:manipulating,bar-tov-tri:j:who-won,bar-tov-tri:j:control} on computational social 
choice.
The various control attacks model different attempts to affect the outcome of 
elections by changes to their structure, namely, via adding, deleting, or 
partitioning 
voters or candidates. The control attack types also vary as to 
whether the goal is to make the focus candidate a winner (perhaps tied; this
is known as the nonunique-winner (NUW) or cowinner setting), 
or to be a 
winner who is not tied  with anyone else (this is known as the unique-winner 
(UW) setting), 
or to \emph{not} (again regarding one of those two variants regarding ties) win.  
Over time, there has been something of a race or contest in
computational social choice to find election systems for which a very large
number of control types have the property that it is $\np$-hard to determine
on a given input whether that type of control can succeed (see the discussion 
in our Related Work section).

But what if two (compatible, i.e., having the same input type) control types 
were in fact the same? That is, what if for every election system (whose votes 
are 
linear orders over the candidates) and on every input, 
despite the control types seeming to model different actions, in fact either for both control types the 
answer is yes
(i.e., for each of the two control types, there is a way to 
achieve the goal of making the focus candidate win (or lose)), or for both 
the 
answer is 
no (i.e., for each of the two control types, there is no way to 
achieve the goal of making the focused candidate win (or lose)). 
We will say that the types ``collapse'' in this case. Then the two types
\emph{for all practical purposes} are the same. In such a case, all research that 
separately studied the two types---for example to determine their computational 
complexity for some election system (over linear orders)---would be doing the same work twice. As our 
abstract mentioned, surprisingly, it has recently been observed by
\citeA{hem-hem-men:j:search-versus-decision} that there are (at least) seven
such collapsing pairs of control types. In fact, that paper shows that four 
control types 
pairwise collapse---yielding six collapsing pairs---and one other pair also 
collapses.

The natural question this raises is whether there are other undiscovered 
collapses---either ones that hold for all election systems or, failing that, 
ones that hold when our focus is limited to one of the most important 
election systems. 
The universe of control types we study
is that of the 22 ``standard'' control types used by~\citeA{hem-hem-men:j:search-versus-decision} 
and other papers (and these are exactly the same 22 control
types that appear in the key table summarizing control
research 
as found in the Handbook of Computational Social Choice, see~\citeR{fal-rot:b:handbook-comsoc-control-and-bribery}), each 
studied in both the NUW and the UW winner model; so 44 control types in total.
For this control-type universe, we completely resolve whether
additional collapses exist, both as to whether any 
additional collapses beyond the seven hold in general---we show that the 
remaining 
hundreds of %
pairs all separate---and as to the major election systems plurality, veto, and
approval, for two of which we do discover additional collapses
(one additional collapsing pair for veto and 14 additional collapsing pairs for approval)
and for all
three of which we then 
separate all remaining pairs. 

Why is this important? As pointed out above, collapses reveal that seemingly different control types 
are for all practical purposes the same ones in disguise, and so general collapses give a 
clearer picture of the world of control types, and can help us avoid 
duplicate work, as has indeed already occurred.\footnote{%
Such duplicate work has already occurred extensively. Each time a paper 
built polynomial-time algorithms for both elements of a collapsing pair
of control types, or proved $\np$-completeness for both elements,
the paper did needless work on one element of the pair, since the sets involved
in such a pair are the same set and so perforce they are of 
identical complexity. As just a few of the many papers that would have been saved a bit
or a lot of work by either the seven general collapses
of 
\citeA{hem-hem-men:j:search-versus-decision}
or the additional concrete-system collapses 
established in the
present paper, we mention~\citeA{fal-hem-hem-rot:j:llull}, \citeA{hem-hem-rot:j:destructive-control}, \citeA{men:j:range-voting}, and~\citeA{nev-rot:c:online-offline-borda-control}.} Separations on the other hand assure us that the types differ, and so we are 
not in the above way 
doing duplicate work in studying 
separately the two control types. (Of course, even for differing 
control types there may be connections of various sorts between the 
two types, e.g., if both are NP-complete, they will
many-one polynomial-time interreduce.)
Since (general-case) collapse
is defined as universal over election systems, its negation
(separation) is existential over election systems,
and as 
this paper 
shows, specific systems may have additional collapses. Thus 
it is important for the field, as to its understanding of control types,
to find the collapse/separation behavior for 
important election systems, and we completely resolve this for 
the three arguably most important ones: plurality, veto, and approval. 
We hope that future papers by others will study additional systems or,
even better, 
will
(as we do for some cases in Section~\ref{sub:av})
define sufficient
conditions that in one fell swoop provide collapses that apply to 
groups of election systems.

Separations themselves can be refined, for two (compatible) control types, 
if on each input (except with no focus candidate specified in the input) we 
consider for each of the two control types the set of all candidates $c$ such that
if $c$ is the focal candidate, the given control action can succeed.  It might 
be the case that for the first control type on all inputs that set is a subset 
of that set for the second control type; or it could always be a superset; or 
neither of these cases could hold. Of course, if the first two cases hold 
simultaneously, that is the same as the two types collapsing. For \emph{all} 
our control pairs that separate, we refine our separations into the three cases 
just described: ``always $\subseteq$, and on some instance $\subsetneq$'';
``always $\supseteq$, and on some instance $\supsetneq$''; and neither of the 
above (i.e., ``incomparable''). We find, for various separating pairs, such 
containment relationships, and not all are UW and NUW variants of each other
(though UW and NUW always have an obvious $\subseteq$ or $\supseteq$ relation; which 
one of those holds depends on whether the type is ``constructive'' ($\subseteq$ holds) or 
``destructive'' ($\supseteq$ holds); we will refer back to this fact later as $(**)$). 

We draw on a wide variety of approaches to establish our separations and collapses. 
Many of our separations are established by human-constructed 
example; since the number of pairs we separate is large, when we can we build examples that simultaneously yield many separations. 
Some of our separation constructions are obtained by computer-aided search. Some of our separation results on the ``general-case'' tables are obtained by inheritance from a specific-case table. As to our new collapses, some are achieved by direct arguments that exploit some feature of the setting, and most are proven via an axiomatic-sufficient-condition approach. Also, some of the collapse (equality) entries in our tables for specific election systems are inherited from the general-case tables. 
Some of our collapses 
draw on multiple approaches. For example, our proof of Theorem~\ref{t:av-cc-c-tp-nuw} 
is a direct proof based on aspects of the setting but also uses an axiomatic property (immunity).

Big picture, this paper completely resolves the extent to which control-type 
pairs collapse or fail to collapse, both in general (i.e., universally 
quantified over election systems) and with regard to plurality, veto, and 
approval elections.

\section{Preliminaries}

For sets $A$ and $B$, $A-B$ denotes $A \cap \overline{B}$, i.e., ``$-$'' denotes the set difference operator.

For consistency, some of the standard definitions that appear in this section are 
taken, at times verbatim, from
\citeA{hem-hem-men:j:search-versus-decision}.

An election comprises a finite set $C$ of candidates (each identified uniquely by a name%
\footnote{\label{f:names}%
By allowing names 
we potentially allow the names
to be 
nefariously 
exploited by election systems.  
However, that model as to the use of names in fact
makes our 
collapses and containments 
\emph{stronger} 
than if those results were in a model where \mbox{(candidate-)neutrality}
is assumed/required (the same applies to
all collapses from 
\citeA{hem-hem-men:j:search-versus-decision}, 
as those results are in this same with-names model).  Although use 
of names would make our separations weaker,
we address that by having ensured that 
every separation we establish in
this paper is achieved via a 
\mbox{(candidate-)neutral} election system.
In contrast, the one separation proven in 
\citeA{hem-hem-men:j:search-versus-decision} 
uses a system that is not 
\mbox{(candidate-)neutral}, see the discussion in our
Related Work section.
}%
) and a
finite 
collection $V$ of votes over $C$. 
Except in one section of the paper (Section~\ref{sub:av}, where we will study
a system using a different vote type, known as approval vectors), we throughout this paper take the ``type'' of a vote as being a
linear ordering over $C$ (note: linear orderings---complete, transitive, antisymmetric binary relations---are inherently tie-free).
A simple example of a 3-candidate, 4-voter election thus is
$C = \{a,b,c\}$ and 
the votes being $a>b>c$, $a>b>c$, $a>c>b$, and $b>c>a$.\footnote{%
The treatment of $V$ as a multiset---which we use to 
match 
\citeS{hem-hem-men:j:search-versus-decision} models---in effect limits the election 
systems the model covers to so-called (voter-)anonymous election systems.
In fact, all general-case collapses of 
\citeA{hem-hem-men:j:search-versus-decision}
and all new general-case containments of our paper also hold
(in the natural nonanonymous model) for all nonanonymous election systems,
as one can easily see from those results' proofs. 
Also, we mention that all three concrete election
systems that we study in fact \emph{are} anonymous.}

As is standard in computational social choice, an election system $\cale$ maps
an election (a pair $(C, V)$) to a (possibly nonstrict) subset of $C$ (the set of winners). As is 
often done in computational social choice papers, we do not 
forbid the case of having no winners;
see~\citeA[Footnote~3]{hem-hem-men:j:search-versus-decision} 
for a discussion of 
why allowing that is 
natural.
The specific election systems plurality, veto, and approval are widely known 
and studied; 
we will explicitly define each in the results 
section devoted to it.

The study of electoral control from a computational perspective was initiated by 
\citeA{bar-tov-tri:j:control}. Their notion, known as constructive control, focuses on making a particular candidate a winner
by some ``control'' action.   
\citeA{hem-hem-rot:j:destructive-control} define the natural variants of those where instead the goal
is to \emph{prevent} a particular candidate from winning. This is 
known as destructive control.

The control actions are: partition of voters, partition of candidates, run-off partition of candidates, deleting voters, deleting candidates, adding voters, and adding candidates (both limited and unlimited).
For each partition-based control type, we have two subvariants to handle the outcome of the subelections: in the TE
(ties eliminate) subvariant, a candidate proceeds to the final round exactly if that candidate is the
unique winner of that subelection, and in the TP (ties promote) subvariant, every winner of a subelection
(including nonunique winners) proceeds to the final round. 
\citeA{hem-hem-men:j:search-versus-decision}
suggest
that it 
is more natural to study the TE subvariant when dealing with the
unique-winner model and the TP subvariant when dealing with the nonunique-winner model. In this paper,
since our goal is to uncover all possible collapses and separations (within the standard control types and their variants), we consider both TE and TP subvariants, regardless of the winner model. Since the nonpartition types among the standard 
control types do not involve two-stage elections, for those we of course will
not split them into TE and TP versions, since that would not make sense.

We take the following definition from~\citeA{hem-hem-men:j:search-versus-decision} (which
itself is taking parts from earlier papers) to help us define the
(constructive) control problems (in the nonunique-winner model)
succinctly.\footnote{%
Regarding %
the partition-based control types,
in 
Definition~\ref{d:control-types} and elsewhere in the paper we sometimes 
will speak of elections
whose candidate set is $C'$ but whose votes, due to candidate partitioning and/or
first-round candidate eliminations, are over a set $C \supseteq C'$.
As is standard in the literature, in such cases we always take this to mean that the
votes are each 
restricted to just the candidates in $C'$.
}

\begin{definition}[see~\citeR{hem-hem-men:j:search-versus-decision}]\label{d:control-types}  %
Let $\cale$ be an election system.
\begin{enumerate}
    \item\label{ccac} In the \textbf{constructive control by adding candidates} problem for $\cale$
    (denoted by $\caledash\ccacnuw$), we are given two
    disjoint sets of candidates $C$ and $A$, $V$ a collection of votes over $C \cup A$, a
    candidate $p \in C$, and a nonnegative integer $k$. We ask if there is a set
    $A' \subseteq A$ such that (i)~$\|A'\| \leq k$ and (ii)~$p$ is a winner of $\cale$ election
    $(C\cup A', V)$.
    
    \item\label{ccacu} In the \textbf{constructive control by unlimited adding candidates} problem for $\cale$ (denoted by $\caledash\ccuacnuw$), we are given two
    disjoint sets of candidates $C$ and $A$, $V$ a collection of votes over $C \cup A$, and a
    candidate $p \in C$. We ask if there is a set
    $A' \subseteq A$ such that $p$ is a winner of $\cale$ election $(C\cup A', V)$.\footnote{In
    Definition~\ref{d:control-types},  
 parts~\ref{ccac} and \ref{ccdc}--\ref{ccdv}
    each have among their inputs
    a
    parameter~$k$ that limits the 
    number of additions or deletions.
    Among those four parts, why does 
    only part~\ref{ccac} 
    have an ``unlimited'' variant
    (part~\ref{ccacu} of the definition)
    included in the standard set of
    control types?  The answer 
    comes from the history of control.  The seminal paper on
    control~\cite{bar-tov-tri:j:control} defined addition/deletion
    of voters and deletion of candidates with a limiting parameter~$k$, but in a curious 
    asymmetry---perhaps due to 
    which version the authors had obtained results for---defined and studied 
    adding candidates in the 
    unlimited variant.  Later papers 
    (the first of them
    being~\citeR{hem-hem-rot:j:destructive-control}) 
ended the omission
by studying 
    the version of adding candidates
    that had a limiting parameter~$k$.
    However, most papers---including the papers this paper is most closely related to---broadly 
    investigating or ``score-carding'' properties 
    of a ``standard'' set of control types have, in addition to studying that with-limiting-parameter-$k$ version of adding candidates, also retained and studied the unlimited version. They likely did that in part since the unlimited version 
    had been introduced in the seminal 
    paper on control, and in part to support comparisons
    with other papers
    that included the unlimited version.}

    \item\label{ccdc} In the \textbf{constructive control by deleting candidates} problem for $\cale$ (denoted by $\caledash\ccdcnuw$), we are given an election
    $(C, V)$, a candidate $p \in C$, and a nonnegative integer $k$. We ask if there is a set
    $C' \subseteq C$ such that (i)~$\|C'\| \leq k$, (ii)~$p \not\in C'$, and
    (iii)~$p$ is a winner of $\cale$ election
    $(C - C', V)$.
    
    \item\label{ccav} In the \textbf{constructive control by adding voters} problem for $\cale$
    (denoted by $\caledash\ccavnuw$), we are given a set of candidates $C$,
    two collections of votes, $V$ and $W$, over $C$, a candidate $p \in C$, and a nonnegative
    integer $k$. We ask if there is a 
    collection $W' \subseteq W$ such that (i)~$\|W'\| \leq k$ and (ii)~$p$ is a winner of $\cale$ election $(C, V \cup W')$.
    
    \item\label{ccdv} In the \textbf{constructive control by deleting voters} problem for $\cale$
    (denoted by $\caledash\ccdvnuw$), we are given an election
    $(C, V)$, a candidate $p \in C$, and a nonnegative integer $k$. We ask if there is a collection
    $V' \subseteq V$ such that (i)~$\|V'\| \leq k$ and (ii)~$p$ is a winner of $\cale$ election
    $(C, V-V')$.
    
    \item In the \textbf{constructive control by partition of voters} problem for $\cale$, in the TP or TE tie-handling rule model
    (denoted by $\caledash\ccpvtpnuw$ or $\caledash\ccpvtenuw$, respectively), 
    we are given an election $(C, V)$, and a candidate $p \in C$. We ask if there is a 
    partition\footnote{%
        A partition of a 
        collection $V$ is a pair of 
        collections $V_1$ and $V_2$ such that $V_1 \cup V_2 = V$, where $\cup$ denotes multiset union.
        A partition of a set $C$ is a pair of sets $C_1$ and $C_2$ such that $C_1 \cup C_2 = C$ and
        $C_1 \cap C_2 = \emptyset$, where $\cup$ and $\cap$ are standard set union and intersection.}
    of $V$ into $V_1$ and $V_2$ such that $p$ is a winner of the two-stage election where the winners of
   subelection $(C, V_1)$ that survive the tie-handling rule compete 
   (with respect to vote collection $V$)
    along with
    the winners of subelection $(C, V_2)$ that survive the tie-handling rule.
    Each election (in both stages) is conducted using election system $\cale$.
    
    \item In the \textbf{constructive control by run-off partition of candidates} problem for $\cale$, in the TP or TE tie-handling rule model
    (denoted by $\caledash\ccrpctpnuw$ or $\caledash\ccrpctenuw$, respectively), 
    we are given an election $(C, V)$, and a candidate $p \in C$. We ask if there is a 
    partition of $C$ into $C_1$ and $C_2$ such that $p$ is a winner of the two-stage
    election where the winners of
    subelection $(C_1, V)$ that survive the tie-handling
    rule compete 
    (with respect to vote collection $V$)
    against the winners of 
    subelection $(C_2, V)$
    that survive the tie-handling rule.
    Each election (in both stages) is conducted using election system $\cale$.
    
    \item In the \textbf{constructive control by partition of candidates} problem for $\cale$, in the TP or TE tie-handling rule model
    (denoted by $\caledash\ccpctpnuw$ or $\caledash\ccpctenuw$, respectively), 
    we are given an election $(C, V)$, and a candidate $p \in C$. We ask if there is a 
    partition of $C$ into $C_1$ and $C_2$ such that $p$ is a winner of the two-stage
    election where the winners of
    subelection $(C_1, V)$ that survive the tie-handling
    rule compete (with respect to vote collection $V$)
    against 
    all candidates in $C_2$.
    Each election (in both stages) is conducted using election system $\cale$.
    \end{enumerate}
\end{definition}

There are 11 control ``types'' listed above (applied regarding generic 
election system $\cale$).  
For each, we can change ``is a winner'' to ``is an untied (i.e., unique) winner'';
for those 11 variants, we change the ``-\nuw'' into ``-\uw.''  Thus we have 22 control types.
Those 22 are all trying to make
the focus candidate win.  And so they are all spoken of as ``constructive'' control types 
(thus the ``CC'' in their naming strings).
Finally, for each of those now 22 control types, we can ask whether 
one can ensure that the focus candidate is \emph{not} a winner or \emph{not} a 
unique winner; those are known as the ``destructive'' control variants,
and in the names of those, the \cc\  is replaced by a \dc\@,
e.g., $\cale$-DC-AC-UW\@.

We thus have 44 total types of control, which we will view as the
``standard'' control types.  We thus have $\binom{44}{2} = 946$ pairs of control 
types. Fortunately, 624 of those pairs are incompatible---the two control types have 
different input fields from each other\footnote{So for example 
\ccpctpnuw\ and \ccrpcteuw\ are compatible, but 
\ccpctpnuw\ and \ccavnuw\ are not, since \ccavnuw\ has a
nonnegative integer field $k$. Throughout this paper, whenever 
we speak of pairs of control types, we refer only to 
compatible pairs (even if that is not explicitly stated, although for emphasis we often do state~it). 
} 
and so comparing them would not 
even make sense---and so we will exclude them from our study.  So we have ``only'' 322 
pairs to focus on in our study. Table~\ref{table:compatibility} shows the five compatibility 
equivalence classes that the 44 types partition into.

When speaking of
control types we often will be speaking of the control model itself,
and when doing so, we generally do not include the ``$\caledash$''
prefix.  In some sense, we view, for example, $\ccacnuw$ as a control
type (model)---one among the 44 standard such models---and 
\emph{we view $\caledash\ccacnuw$ as the set of input strings
that are ``yes'' instances, for
election system $\cale$, under that model of control}.  

Keeping that in mind, we now define precisely what we mean by two (compatible) control types collapsing or 
separating, and then introduce a function approach that will allow us to explore
in a more refined way the nature of the separations.
Let $\cale$
(e.g., Plurality)
be an election system and let $\calt$ and $\calt'$ be two (compatible) control 
types from our 44 standard ones
(e.g., CC-AC-NUW and CC-AC-UW).  Then if ${\caledt} = {\caledt'}$  we say 
that the  control types $\calt$ and $\calt'$ collapse (for election system $\cale$), 
and if  ${\caledt} \neq {\caledt'}$  we say 
that the control types $\calt$ and $\calt'$ separate (for election system $\cale$).
We will also use the terms collapse and separate for a more general case, namely,
the one of quantifying universally over all election systems whose votes 
are 
linear orders.  When speaking in that case, we will say that two (compatible)
control types from among our 44 collapse if the two control types collapse for every 
election system $\cale$ whose vote type is 
linear orders, and otherwise 
we will say that the control types separate.

As an example, if we fix the election system $\mathcal{E}$ to be plurality, we note that in the
3-candidate, 4-voter election example given at the beginning of this section $a$ is a unique winner.
Yet, we could make $b$ a unique winner by deleting candidate $a$, hence 
we have
$(C,V,b,1) \in\textnormal{Plurality-CC-DC-UW}$.
Also, since a unique winner is a winner in the nonunique-winner model as well, we have $(C,V,b,1)\in\textnormal{Plurality-CC-DC-NUW}$. (In this 
paper we will not focus on encoding details,
since they are not important to our study.) 

Note that, in this paper, when we 
prove that two compatible control types separate we are unconditionally
proving that they are not the same set. That contrasts with 
the 
weaker---namely, conditional---separation 
one might claim if one of the types was known to be NP-hard 
and the other was known to be a set (other than the empty set 
or $\Sigma^*$) in P, since arguing separation just from those facts 
would
require
one to have proven that $\pneqnp$. Also, one needs to be wary
about casually arguing that certain (compatible) control types ``inherently,''  
universally
differ. For example,
it might be tempting to say just based on intuition that 
CC-DV-UW and CC-DC-NUW inherently differ for all election systems.
However, for the election system where all candidates always lose,
these two do collapse, since both are the empty set. Also, the already-known
universally collapsing control pairs found by 
\citeA{hem-hem-men:j:search-versus-decision}
were long assumed to be obviously different, but that turned 
out to not be the case. (However, though in this paper we do not 
look at showing when pairs of compatible types will differ for 
every election system, we mention that in some cases one can indeed
see that that holds. For example, for each election 
system $\cale$, $\cale$-CC-DV-NUW and 
$\cale$-DC-DV-NUW differ since any given legal, syntactically well-formed
input with one candidate and 
zero votes, that input will be in exactly one of the two sets.)

Compatible control types that separate---i.e., are different sets---can do so in different ways, some 
of which reflect containment relationships.  In order to be able to 
seek such relationships, we define the three different ways that 
two (compatible) separating control types, $\calt$ and 
$\caltprime$, for $\cale$ can separate.  To support this refinement, 
we introduce a function model.  In particular, we will define 
functions that, for a given control type, map from inputs 
(that differ from  
those used so far for that control type only in \emph{not} having
a focus candidate specified) to the set of all 
candidates $c$ such that if $c$ is made the focus candidate 
for that input, successful control is possible.
In a bit more detail: 
Each of our control types has certain inputs, and all include a 
focus candidate, $c$.  Let us for
any of our 44 control types, $\calt$, refer to an input 
to it, except with the field containing the focus candidate removed,
as the reduced form of that input. If we want for 
a reduced input to add back
the name of a particular candidate to be the focus candidate, 
we will say that is inflating the reduced input by that candidate.
For any of our 44 control types, $\calt$, 
and any election system $\cale$, we define the function ${f}_{\caledt}$
to be the function that, for a given reduced input $\tilde{I}$, outputs the 
set of all candidates $c$ such that $\tilde{I}$ inflated by $c$ 
belongs to the set $\caledt$.  For example, given as input $(C,V,3)$, the output of $f_{\caledash\dcdvnuw}$ is the set of 
all candidates $c$ that can by deleting less than or equal to three 
votes be prevented from being a winner. 

Clearly, two (compatible) control types collapse exactly if 
their thus-defined functions are equal.  But for two separating (compatible)
control types, there are three different ways they can 
be separated.
One way is if, for each reduced input $\tilde{I}$:
(a)~$f_{\caledt}(\itilde) \subseteq f_{\caledt'}(\itilde)$ and (b)~for some reduced input $\itilde'$,
   $f_{\caledt'}(\itilde') - f_{\caledt}(\itilde') \neq \emptyset$.
We will (in a slight abuse of notation)
refer to the case where~(a) holds as the ``$\subseteq$'' case,
and the case where both~(a) and~(b) hold as the ``$\subsetneq$'' case.
The ``$\supseteq$'' and ``$\supsetneq$'' cases 
are analogously defined.
If the two (compatible) types separate but neither the
``$\subsetneq$'' case nor
the ``$\supsetneq$'' case holds,
we will say the two types are incomparable:
each will sometimes
have successful focus candidates that the other does not.
If both directions of noncontainment can be witnessed by 
the same reduced input, we will say the two types are strongly incomparable,
i.e., there is a reduced input $\itilde$ such that 
$f_{\caledt}(\itilde) - f_{\caledt'}(\itilde) \neq \emptyset$
and
$f_{\caledt'}(\itilde) - f_{\caledt}(\itilde) \neq \emptyset$.
Arguably, strong incomparability is a more satisfying strength 
of incomparability, since it makes clear that 
one does not have to cobble the two directions of the incomparability together from different supporting instances.
These notions are relative to each specific election system, $\cale$.

For the general case---where our collapses are universally
quantified over all election systems whose vote type is 
linear orders---we define three increasingly strong 
types of incomparability.  
The weakest incomparability notion
is simply that for at least one such election system
$\cale$,
for some reduced input $\itilde$,
$f_{\caledt}(\itilde) - f_{\caledt'}(\itilde) \neq \emptyset$ holds, and for at least one 
such election system $\cale'$, 
for some reduced input $\itilde'$,
$f_{\cale'\hbox{-}\calt'}(\itilde') - f_{\cale'\hbox{-}\calt}(\itilde') \neq \emptyset$
holds.
We will call this being weakly incomparable.\footnote{We will state no weak incomparability 
results in this paper, since whenever we obtained a weak 
incomparability result we in fact were able to even establish  
incomparability, which we will 
define in a moment in the main text.
Though in this paper we thus never need to prove 
weak incomparability 
results, we state the notion so that 
future research on other control types will have a 
three-level menu of incomparability strengths to draw on if 
needed.}
We will say the pair is 
incomparable if there is some election system, over 
linear orders,
in which the pair is incomparable in the sense of
the previous paragraph. 
And we will say the pair is 
strongly incomparable if there is some  election system, over 
linear orders,
in which the pair is strongly incomparable in the sense of 
the previous paragraph. 
For the general case, we will say ``$\subseteq$'' 
(resp.,\ ``$\supseteq$'') holds 
if for every election system $\cale$ whose vote type is 
linear orders, the ``$\subseteq$'' 
(resp.,\ ``$\supseteq$'') case for $\cale$, as defined 
above, holds.

We now cover, and give brief reference labels to (though we at 
times may use these inheritances tacitly when the use is clear), 
some collapse, containment, and separation inheritances
that hold.  
We will start these labels with the letter ``I'' as a mnemonic
for that fact that these concepts are about inheritance.
General-case collapse, $\subseteq$, and 
$\supseteq$ results imply, for each
election system over 
linear orders, 
resp.,\ collapse, $\subseteq$, and
$\supseteq$ results (let us shorthand that fact as I1).
If a given general case result of this type 
\emph{happens} to in addition hold for 
every election system (not merely those over 
linear orders), we will refer to that as additionally
being an I1${}^+$ case\@.
Simply because their particular proofs do not ever draw
on the vote types, the seven general-case collapses of~\citeA{hem-hem-men:j:search-versus-decision},
and also our general containment results of
Theorem~\ref{t:general-containments}, hold even as 
I1${}^+$ cases. 
As to separation inheritances, 
if for some election system $\cale$ over 
linear orders and some reduced input $\itilde$ 
we have that 
$f_{\caledt'}(\itilde) - f_{\caledt}(\itilde) \neq \emptyset$
(as, crucially, is always the case if we have that for $\cale$
the $\subsetneq$ relation holds between $\calt$ and $\calt'$)
and we in addition happen to have that in the general
case the $\subseteq$ relation is known to hold 
between $\calt$ and $\calt'$, then we may conclude
that $\subsetneq$ holds in our general case; the 
analogous claim holds for the $\supsetneq$ case, and we 
will refer to either of these, when they hold, as 
an I2 inheritance case.  
Incomparability in an election 
system (over 
linear orders) implies general-case 
incomparability (we will refer to this as I3), and strong incomparability in an election
system (over 
linear orders) implies general-case 
strong incomparability (I3${}^*$).

\section{Results%
}

In the following three subsections, for the general case and for the
cases of plurality, veto, and approval voting, we completely determine which
pairs of (compatible) standard control types collapse, and which separate.
Beyond that, we refine every separation into one of the
three disjoint cases: $\subsetneq$, $\supsetneq$,
and incomparability.   

We discover a 
number of previously unknown collapses (for two of
the three specific systems) and also for many noncollapsing control-type pairs 
establish new containments in one direction.
Regarding the former, for veto we discover a new collapsing pair
$(\vetodcpvtenuw=\vetodcpvteuw)$ and for approval we extend a four-type
collapse by 
\citeA{hem-hem-men:j:search-versus-decision}
to a six-type
collapse, and we find five additional collapsing pairs, for a total
of $\binom{6}{2} + 5 - \binom{4}{2}$, i.e., $14$, new collapsing control-type pairs for approval. 
Overall, we establish that, for the $4 \times 322 = 1{,}288$ cases we study
(i.e., all compatible pairs for each of the four cases),
no containments or collapses exist other than those provided by either 
\citeA{hem-hem-men:j:search-versus-decision} or this paper.

\subsection{The General Case and Plurality}\label{sub:general}

We show that the only (compatible) pairs that collapse
in general (i.e., that collapse for every election system)
are the seven found by 
\citeA{hem-hem-men:j:search-versus-decision}, namely,
for every election system $\cale$,
$\caledash\dcrpctenuw=\caledash\dcrpcteuw=\caledash\dcpctenuw=\caledash\dcpcteuw$ (six pairs) and $\caledash\dcrpctpnuw=\caledash\dcpctpnuw$ (one pair).
Surprisingly, we will be able to do so using just constructions about plurality elections,
combined with uses of inheritance.

In a plurality election,
for each vote where a particular candidate is ranked first, that candidate
receives a point, and a winner is a candidate with the highest number
of points among all the candidates (naturally, there can be multiple
winners). 
The votes in plurality elections are 
linear orders. Recall that by the
``general case,'' we mean the general case with respect to elections where the votes are
linear orders. Thus to separate two control types in the general case, it certainly suffices
to show that they separate under plurality, as such separations inherit back to the general case
via the I3 and I2 
types of inheritance discussed in
our preliminaries.\footnote{We note that for some cases
we achieve strong
incomparability in our general-case results, always via I3${}^*$. 
In such cases, incomparability still holds, since strong incomparability is simply a specific subcase of 
incomparability.}
The reason we know that no separation for the general case is missed is that our results show that
plurality has 
no $\subseteq$, $\supseteq$, 
or collapse results not also possessed by the general case;
so our I2 cases are valid, 
and every general-case $\subsetneq$, $\supsetneq$, or incomparability
(or even strong incomparability) that holds is yielded by our inheritances.

Let us now turn our attention to the two new general-case containments shown in this paper (these containments are in addition to the obvious ones, noted at location $(**)$ of Section~\ref{sec:introduction},
about two control types that only differ in their
winner model) and also argue that both are strict in the general case (by which, recall, we 
mean that there is at least one election system under which the containment is not an equality).
The following two containments apply
to all vote types (not just 
linear orders; and we argue in Section~\ref{sub:av} that the collapses by 
\citeR{hem-hem-men:j:search-versus-decision}
also apply to all vote types).

\begin{theorem}\label{t:general-containments} %
    Let $\cale$ be an election system.
    For each $\calt \in \{\dcrpctpuw, \dcpctpuw\}$, 
    $\caledt \subseteq \caledash\dcrpctenuw$.
\end{theorem}
\begin{proof}
    Let $\calt \in \{\dcrpctpuw, \dcpctpuw\}$.
    We will show $\caledt \subseteq \caledash\dcrpctenuw$.  
    Suppose $(C,V,p) \in \caledt$.
    Let $(C_1, C_2)$ be a partition that witnesses $(C, V, p)$'s membership in $\caledt$.
    It holds that either $p$ participates in and loses in a subelection, or
    $p$ participates in and does not win uniquely in the final round.
    If the former holds, then the partition $(C_1, C_2)$ suffices as $p$ will be
    eliminated in a subelection and thus will not proceed to the final round.
    If the latter holds, then let $D$ denote the set of candidates present 
    during the final round. Clearly $p\in D$.
    Thus the partition $(D, C-D)$ witnesses $(C, V, p)$'s membership in $\caledash\dcrpctenuw$ since
    $p$ either loses in the first subelection $(D, V)$ or ties in that subelection and is
    eliminated by the tie-handling rule.~\end{proof}

For both containments above
(i.e., for each of the two values of $\calt$ stated in the theorem), Table~\ref{table:plurality-results} contains a pointer to a separation witness 
in Table~\ref{table:plurality-tools} that shows that the containment is not an equality in the general case (i.e., that for some election system---in fact, plurality---the containment is strict). More generally, Table~\ref{table:plurality-results}, for each compatible pair of 
control types $\calt$ and $\caltprime$, gives us an election witnessing 
$\pluralitydash\calt - \pluralitydash\caltprime \neq \emptyset$ if that holds, and gives an
election witnessing $\pluralitydash\caltprime - \pluralitydash\calt \neq \emptyset$
if that holds (and if both hold Table~\ref{table:plurality-results} gives witnesses for each). Of course, we are not claiming that the containments in that 
theorem are strict under all election systems; for example, 
Corollary~\ref{c:approval-axiomatic-results} and Theorem~\ref{t:approval-extend-to-6-collapse} show that all three types mentioned in the above theorem pairwise collapse under
approval voting.

Some of the constructions in Table~\ref{table:plurality-tools} are quite simple.
For example, with $C = \{a, b\}$ and $V = \{a>b\}$ (denoted by ``Plur.3'')
we clearly get incomparability between all 144 
pairs 
of partition types where one type is constructive and the other is destructive. 
(This holds since in that election, under every constructive
or destructive partition-control action, $a$ is the unique winner
and $b$ is not a winner.)
However, showing separations for every
pair that can be separated is no trivial matter. Some of our
separation examples for plurality are quite large, with
up to 18 votes 
and up to seven candidates. 
The search for those examples was computer-aided and extensive.

\subsection{Veto}\label{sub:veto}

In a veto election, for each vote where a particular candidate is not ranked last, that candidate
receives a point, and a winner is a candidate with the highest number
of points among all the candidates (naturally, there can be multiple
winners).
For example,
if $C = \{a, b, c\}$ and $V = \{a>b>c, c>a>b\}$, then $b$ and $c$ each
receive one point, and $a$ receives two points and wins.

We now establish every equality and containment that holds for veto elections but was not 
established by one of 
the results of 
\citeA{hem-hem-men:j:search-versus-decision}.  For all other veto cases,
we have constructed counterexamples.
Some of those counterexamples
were obtained through computer search.

\begin{theorem}\label{t:veto-new-collapse} %
   $\vetodcpvteuw = \vetodcpvtenuw$.
\end{theorem}
\begin{proof}
    The $\supseteq$ relationship is immediate.
    The approach that follows resembles closely that of 
    \citeA{mau-rot:j:control-veto-plurality}.
    Let $(C, V, p) \in \vetodcpvteuw$. If there is a 
    partition such that $p$ is not a winner
    of the two-stage election, then $(C, V, p)
    \in \vetodcpvtenuw$. Otherwise, there is a partition
    and $c \in C$, $c\neq p$, such that $p$ and $c$ are both
    winners of the two-stage election.
    Suppose $\|C\| = 2$. Then $p$ and $c$ are both winners of the
    election $(C, V)$. In this case,
    consider the partition $(V, \emptyset)$. Both candidates will tie and be eliminated, in both subelections.
    Suppose $\|C\| \geq 3$. Then there are two distinct
    candidates $d,e \in C-\{p\}$.
    Let $V_1$ denote the set of votes in which $e$ is vetoed and let $V_2$ be the remaining votes. Now, consider
    the two-stage election with partition $(V_1, V_2)$. 
    Since $e$ is never vetoed in $V_2$, $p$ can at best tie with
    $e$ in the subelection $(C, V_2)$. Finally, $d$ is never
    vetoed in $V_1$ (since all votes there veto $e$), so $p$ cannot be a unique winner of
    $(C, V_1)$.  Thus $p$ is eliminated in both first-round elections.~\end{proof}

\begin{theorem}\label{t:veto-new-containments-pv} %
    For each $\calt \in \{\dcpvtpuw$, $\dcpvtpnuw\}$, $\vetodash\calt \subsetneq \vetodcpvtenuw$.
\end{theorem}
\begin{proof}
    Let $\calt \in \{\dcpvtpuw$, $\dcpvtpnuw\}$.
    Suppose $(C, V, p) \in \vetodash\calt$,
    and let $(V_1, V_2)$ be a partition that witnesses this membership.
    Using the same technique as in the proof of Theorem~\ref{t:veto-new-collapse}, we can handle the case
    where $\|C\| \geq 3$ and thus we only 
    need to consider the case where $\|C\| = 2$ in this proof. Let $C = \{c, p\}$.
    First, suppose that $p$ is present in the final round.
    Then, since $p$ is not a unique winner of the final round, all candidates in
    $C$ are present in this round.
    Thus $p$ is not a unique winner of the election $(C, V)$,
    and so the partition $(V, \emptyset)$ witnesses $(C, V, p)$'s membership in $\vetodcpvtenuw$.
    Now, suppose that $p$ is absent from the final round.
    Then $p$ is not a winner of $(C, V_1)$ or $(C, V_2)$,
    and so $p$ receives strictly more vetoes than $c$ in both $V_1$ and $V_2$.
    Thus $p$ receives more vetoes than $c$ in $V$,
    so again, $p$ is not a unique winner of $(C, V)$.
    A separation witness for the strict containment
    can be found in Table~\ref{table:veto-results}.~\end{proof}

\begin{theorem}\label{t:veto-new-containments-rpc-pc} %
    For each $\calt \in \{\dcrpctenuw$, $\dcrpctpuw$, $\dcrpctpnuw$, $\dcpctpuw\}$, $\vetodash\calt \subsetneq \vetodcpvtenuw$.
\end{theorem}
\begin{proof}
    Let $\calt \in \{\dcrpctenuw$, $\dcrpctpuw$, $\dcrpctpnuw$, $\dcpctpuw\}$.
    Suppose $(C, V, p) \in \vetodash\calt$,
    and let $(C_1, C_2)$ be a partition that witnesses this membership.
    As per the two preceding proofs
    (namely, of Theorems~\ref{t:veto-new-collapse} and~\ref{t:veto-new-containments-pv}), 
    we need only consider the case where $\|C\| = 2$.
    The case where $p$ is present in the final round is handled as in Theorem~\ref{t:veto-new-containments-pv},
    so suppose that $p$ is absent from the final round.
    Without loss of generality, we may assume that $p \in C_1$.
     (This is certainly true when $\calt$ is about one of the theorem's three RPC cases, due to the symmetric nature of that partitioning action. It also holds when $\calt$ is about the theorem's one PC case, since $p$ being in $C_2$ would imply that $p$ is in the final round, which contradicts the case we are in.)
    Then $p$ is not a unique winner of $(C_1, V)$, so $C_1 = C$.
    Thus $p$ is not a unique winner of $(C, V)$,
    and the partition $(V, \emptyset)$ witnesses $(C, V, p)$'s membership in $\vetodcpvtenuw$.
    A separation witness for the strict containment
    can be found in Table~\ref{table:veto-results}.~\end{proof}

\subsection{Approval Voting}\label{sub:av}

Approval voting differs from the other election systems discussed
so far in this paper as to its vote type.  In an 
approval election $(C, V)$, each vote is a
bit-vector of length $\|C\|$, with each bit being associated
with a candidate. If a candidate's bit is 1, then
that candidate is approved by that vote.
In approval voting, the winner set is composed of each candidate $c$
such that no other candidate is approved by strictly more votes than $c$ is.
We will sometimes speak of the ``score'' of a candidate in the rest of this section. 
In the context of approval voting,
the score of a candidate in an election is the number of votes that approve that
candidate (and as noted above, the set of winners are those candidates with maximal score).

Although the collapses shown by 
\citeA{hem-hem-men:j:search-versus-decision} were
stated for election systems where votes are 
linear 
orders, we note that in their proof they do not use the vote type and thus those collapses hold regardless of 
what type of 
votes the election system is over. Thus we have the following corollary.

\begin{corollary}[see~\citeR{hem-hem-men:j:search-versus-decision}]
\label{cor:av-hem-hem-men-20}
\begin{enumerate}
		\item $\approvaldcrpcteuw=\approvaldcrpctenuw=\approvaldcpctenuw=\approvaldcpcteuw$.
		\item $\approvaldcrpctpnuw = \approvaldcpctpnuw$.
	\end{enumerate}
\end{corollary}

In this section, we prove a number of collapses and inclusions that hold for approval voting.
Some of these results draw on previously established immunity\footnote{\label{f:immunity}In the unique-winner model, we say an election system is immune to a particular type of control
if the given type of control can never 
change a candidate from not uniquely winning to uniquely winning
(if the control type is constructive) or change a candidate from uniquely winning to not uniquely winning (if the control type is destructive) (\citeR{hem-hem-rot:j:destructive-control}, which clarified a slightly flawed immunity definition of~\citeR{bar-tov-tri:j:control}).
In the nonunique-winner model, we say an election system is immune to a particular type of control
if the given type of control can never
change a nonwinner to a winner (if the control type is constructive) or change a winner to a nonwinner (if the control type is destructive).}
arguments that draw on 
properties
satisfied
by approval voting, thereby allowing us to generalize our results. Other results are provided by direct
arguments. 
Our direct arguments often rely on the fact that, since the votes are bit-vectors,
a candidate's score in an election is independent of the other candidates present in the election. We will start by discussing the results that draw on the aforementioned
immunity arguments.

Let us first consider 
what is known in the social choice 
literature as Property~$\alpha$ (\citeR{che:j:rational-selection}, see also~\citeR{%
sen:j:choice-functions}).\footnote{In a number of papers in the computational social choice literature,
this property has  been spoken of as if it were the Weak Axiom of 
Revealed Preference.  It in fact is not, though they are related; this issue however
did not make any of the earlier papers wrong, as they merely were unwisely
but internally consistently using a notation in a way that did not match
the way it was used elsewhere.
This unfortunate terminological confusion
came in due to some ambiguous wording in one of the seminal papers on
computational social choice theory. In the present paper, we follow the lead of
\citeA{car-cha-hem-nar-tal-wel:t5-w-aamas-ptr:search-vs-search} and use
the (more consistent with 
the social choice literature) terminology 
``Property~$\alpha$'' and, 
for Property~$\alpha$'s 
``unique-winner'' analogue,
``Property Unique-$\alpha$.'' 
See 
\citeA{car-cha-hem-nar-tal-wel:t5-w-aamas-ptr:search-vs-search} for a more detailed discussion of 
this issue, including its history and relation to the Weak Axiom of 
Revealed Preference.} 
Property~$\alpha$ is defined as the 
property  
that $p$ winning an election $(C, V)$ implies that $p$ remains a winner of every
election $(C', V)$ for which $p\in C' \subseteq C$.
(Clearly, some election systems will satisfy this property and some will not.)
The ``unique'' version of this property (which is called
Property Unique-$\alpha$) only differs in that
it requires $p$ to be a unique winner, i.e., 
it requires that $p$ uniquely winning in an election $(C, V)$
implies that $p$ uniquely wins in each election $(C', V)$ for which $p \in C' \subseteq C$~\cite{hem-hem-rot:j:destructive-control}.
\citeA{hem-hem-rot:j:destructive-control} note that any election system that satisfies Property Unique-$\alpha$
is immune to several types of control, namely
to (i)~destructive control by partition of candidates and run-off
partition of candidates (under both \tp\ and \te\ tie-handling rules) in the \uw\ model, and
(ii)~destructive control by deleting candidates in the \uw\  model.
From this, we obtain the following results.\footnote{The 
theorem 
certainly could also include  $\caledash\dcrpcteuw$ in the equality.  However, including that in the theorem would make little sense, since 
$\caledash\dcrpcteuw = \caledash\dcpcteuw$ does not rely on Property Unique-$\alpha$, but in fact
holds
for \emph{all} election systems in light of 
\citeA{hem-hem-men:j:search-versus-decision} and our comment in the paragraph before Corollary~\ref{cor:av-hem-hem-men-20}.

Also, note that in our sequence of result statements from Theorem~\ref{t:unique-alpha-collapse} through Corollary~\ref{c:av-cc-uw-containments}, each is (intentionally) stated
as applying to all election systems satisfying the relevant Property $\alpha$ or Unique-$\alpha$, regardless of what vote ``type''---such as approval vectors or linear orders---the election system's votes are over.} 

\begin{theorem}\label{t:unique-alpha-collapse} %
Let $\cale$ be an election system that satisfies Property Unique-$\alpha$\@. Then
    $\caledash\dcpctpuw = \caledash\dcpcteuw$.
\end{theorem}
\begin{proof}
Fix any $\cale$ that satisfies Property Unique-$\alpha$\@.
Let $\caledt_1 = \caledash\dcpctpuw$ and let $\caledt_2 = \caledash\dcpcteuw$.
Consider the  two sets $A_\cale = \{(C, V, p) \condition p \in C$ and
$p$ is 
a unique winner of $\cale$ election $(C, V)\}$ and 
$B_\cale = \{(C, V, p) \condition p \in C$ and
$p$ is not a unique winner of $\cale$ election $(C, V)\}$.
These sets 
form a partition of 
$Y= \{(C,V,p) \condition p\in C$ and $(C,V)$ is an $\cale$ election$\}$, so of course 
$Y = A_\cale \cup B_\cale$.
Clearly we also have that 
$\caledt_1 \subseteq Y$ and that $\caledt_2 \subseteq Y$.
We will argue that $\caledt_1 = B_\cale = \caledt_2$.
Since $\cale$, like all 
systems satisfying Property Unique-$\alpha$, is immune to
both control types 
in the theorem statement, we have (recall that both of these types are destructive types) that 
$\caledt_1 \cap A_\cale = \emptyset$ and  $\caledt_2 \cap A_\cale = \emptyset$, and thus
it holds that $\caledt_1 \subseteq B_\cale$ and $\caledt_2 \subseteq B_\cale$.
Fix $(C, V, p) \in B_\cale$. Then the partition $(\emptyset, C)$ witnesses both
$(C, V, p) \in \caledt_1$ and $(C, V,p) \in \caledt_2$, since in both cases the final round will simply be $(C, V)$ and we know that since $(C, V, p) \in B_\cale$, $p$ will not be a unique winner of the final round.~\end{proof}

\begin{theorem}
\label{t:approval-dc-unique-alpha}
Let $\cale$ be an election system that satisfies Property Unique-$\alpha$\@. %
Then the following hold.
\begin{enumerate}
    \item $\caledash\dcdcuw\subseteq\caledash\dcdvuw$.
    \item $\caledash\dcdcnuw\subseteq\caledash\dcdvuw$.
\end{enumerate}
\end{theorem}
\begin{proof}
Fix any $\cale$ that satisfies Property Unique-$\alpha$\@.
\begin{enumerate}
\item
The proof is analogous to that of Theorem~\ref{t:unique-alpha-collapse}.
Since $\cale$ is immune to destructive control by deleting candidates in the unique-winner model (as per the discussion immediately before Theorem~\ref{t:unique-alpha-collapse}), the inclusion is immediate as 
$\caledash\dcdcuw$ is comprised of those inputs where the focus candidate is already not a unique winner.
\item It trivially holds that $\caledash\dcdcnuw \subseteq \caledash\dcdcuw$. Thus the proof
follows from the previous part of this theorem.\qedhere
\end{enumerate}~\end{proof}
\begin{theorem}\label{t:unique-alpha-dc-nuw-dv-uw}
Let $\cale$ be an election system that satisfies Property~$\alpha$\@. 
    Then $\caledash\dcdcnuw \subseteq \caledash\dcdvnuw$.
\end{theorem}
\begin{proof}
Fix any $\cale$ that satisfies Property~$\alpha$\@.
The proof is analogous to that of Theorem~\ref{t:approval-dc-unique-alpha}.
Given an election $(C, V)$ and $p\in C$, if $p$ is a winner of $\cale$ election
$(C, V)$, by Property~$\alpha$ it follows that no deletion of candidates (other than $p$) can prevent $p$ from being a winner. Thus
$\caledash\dcdcnuw$ is comprised of those inputs where the focus candidate is already not a winner.~\end{proof}

The next two results will 
use the notion of being strongly voiced.
An election system $\cale$ is said to be strongly voiced exactly if for every election $(C, V)$ with $C \neq \emptyset$, there is at least one winner of $(C, V)$ under $\cale$~\cite{hem-hem-rot:j:destructive-control}.
(Most
earlier versions of this paper had different 
hypotheses in what here are Theorem~\ref{t:unique-alpha-cc-44} and Corollary~\ref{c:av-cc-uw-containments} 
and 
flawed proofs of 
those; the current version corrects that.)

\begin{theorem}\label{t:unique-alpha-cc-44} %
	Let $\cale$ be a strongly voiced election system that satisfies Property~$\alpha$.	Then
	$\caledash\ccpctpuw = \caledash\ccrpctpuw$.
\end{theorem}
\begin{proof}
Fix any $\cale$ that is strongly voiced and satisfies Property~$\alpha$.
Let $\caledt_1 = \caledash\ccpctpuw$ and let $\caledt_2 = \caledash\ccrpctpuw$.
Consider the  two sets $A_\cale = \{(C, V, p) \condition p \in C$ and
$p$ is 
a unique winner of $\cale$ election $(C, V)\}$ and 
$B_\cale = \{(C, V, p) \condition p \in C$ and
$p$ is not a unique winner of $\cale$ election $(C, V)\}$.
These sets 
form a partition of 
$Y= \{(C,V,p) \condition p\in C$ and $(C,V)$ is an $\cale$ election$\}$, so of course 
$Y = A_\cale \cup B_\cale$.
Clearly we also have that 
$\caledt_1 \subseteq Y$ and that $\caledt_2 \subseteq Y$.
We will show that $\caledt_1 = A_\cale = \caledt_2$. First we show that $\caledt_1 \subseteq A_\cale$ and $\caledt_2 \subseteq A_\cale$.

Consider an element $(C, V, p) \in \caledt_1$ and assume for the sake of contradiction that $p$ is not a unique winner of $\cale$ election 
$(C, V)$, i.e., $(C, V, p) \in \caledt_1 \cap B_\cale$. 
Since $(C, V, p) \in B_\cale$ 
and since $\cale$ is strongly voiced, there is a candidate $d \in C-\{p\}$ 
that is a winner of $(C, V)$ under $\cale$. 
By Property~$\alpha$, that candidate is a winner of any subelection they participate in (regardless of how the candidates are partitioned), and since the tie-handling rule never eliminates candidates, $d$ will always proceed to the final round and be a winner (again by Property~$\alpha$). This contradicts the fact that $(C, V, p) \in \caledt_1$, so $\caledt_1 \cap B_\cale = \emptyset$.
Since $\caledt_1 \subseteq Y = A_\cale \cup B_\cale$ and since $\caledt_1 \cap B_\cale = \emptyset$, it follows that $\caledt_1 \subseteq A_\cale$.
The same argument establishes that $\caledt_2 \subseteq A_\cale$.
 We will now complete this proof by arguing that $A_\cale \subseteq \caledt_1$ and $A_\cale \subseteq \caledt_2$.

Fix $(C, V, p) \in A_\cale$. Then the partition $(\emptyset, C)$ witnesses that $(C, V, p) \in \caledt_1$ as the final round will simply be $(C, V)$ and we know that since $(C, V, p) \in A_\cale$, $p$ will be a unique winner of the final round.
Additionally, the partition $(\emptyset, C)$ also witnesses 
that $(C, V, p) \in \caledt_2$ as no one will proceed from subelection $(\emptyset, V)$, and only $p$ will proceed from subelection
$(C, V)$ (since $(C, V, p) \in A_\cale$), and $p$ must be the unique winner of the final round $(\{p\}, V)$ (this follows from both strong voicedness and Property~$\alpha$).~\end{proof}
We can build on
Theorem~\ref{t:unique-alpha-cc-44} to show additional containments.

\begin{corollary}\label{c:av-cc-uw-containments} %
    Let $\cale$ be a strongly voiced election system that satisfies Property~$\alpha$ and Property Unique-$\alpha$\@.
    Then, for each $\calt \in \{\ccpcteuw$, $\ccpctenuw$, $\ccrpcteuw$, $\ccrpctenuw$, 
    $\ccpvteuw$, $\ccpvtenuw$, $\ccpvtpuw$, $\ccpvtpnuw\}$,
    it holds that $\caledash\ccpctpuw \subseteq \caledt$ (equivalently, $\caledash\ccrpctpuw \subseteq \caledt$).
\end{corollary}
\begin{proof}
    Fix any $\cale$ that is strongly voiced and that satisfies Property~$\alpha$ and Property Unique-$\alpha$, fix
    a $\calt \in \{\ccpcteuw$, $\ccpctenuw$, $\ccrpcteuw$, $\ccrpctenuw$, 
    $\ccpvteuw$, $\ccpvtenuw$, $\ccpvtpuw$, $\ccpvtpnuw\}$,
    and fix $(C, V, p) \in \caledash\ccpctpuw$.
    By the proof of Theorem~\ref{t:unique-alpha-cc-44}, $\caledash\ccpctpuw$ is
    comprised of only those inputs where $p$ is already a unique winner of the $\cale$
    election $(C, V)$, so
    the trivial partition (i.e., $(\emptyset, C)$ if $\calt$ is about partitioning candidates
    and $(\emptyset, V)$ if $\calt$ is about partitioning votes) is a witness that
    $(C, V, p) \in \caledt$ as $p$ will uniquely win at least one subelection in which it participates (if 
    $\calt$ is about partitioning votes, then $p$ participates in two subelections) as one of those two subelections is $(C, V)$. Moreover, $p$ uniquely wins 
    the final round, which it participates in, because
    $\cale$ satisfies Property Unique-$\alpha$.  The ``equivalently'' follows from Theorem~\ref{t:unique-alpha-cc-44}.~\end{proof}

Since approval voting is clearly strongly voiced~\cite{hem-hem-rot:j:destructive-control} and satisfies
Property Unique-$\alpha$~\cite{hem-hem-rot:j:destructive-control},
and (clearly) Property~$\alpha$,
the above theorems and corollaries apply to approval voting.
For each case where only the containment (and not the collapse) is shown, the containment can
be made strict under approval voting. See Table~\ref{table:approval-results} for the separation witnesses.

\begin{corollary}\label{c:approval-axiomatic-results}
\begin{enumerate}
    \item $\approvaldcpctpuw=\approvaldcpcteuw
    = \approvaldcrpcteuw = \approvaldcrpctenuw = \approvaldcpctenuw$.
    \item $\approvaldcrpctpnuw = \approvaldcpctpnuw$.
    \item $\approvaldcdcuw \subsetneq \approvaldcdvuw$.
    \item $\approvaldcdcnuw \subsetneq \approvaldcdvnuw$.
    \item $\approvaldcdcnuw \subsetneq \approvaldcdvuw$.
    \item $\approvalccpctpuw = \approvalccrpctpuw$.
    \item For each $\calt \in \{\ccpcteuw$, $\ccpctenuw$, $\ccrpcteuw$, $\ccrpctenuw$, 
    $\ccpvteuw$, $\ccpvtenuw$, $\ccpvtpuw$, $\ccpvtpnuw\}$,
    it holds that $\approvalccpctpuw \subsetneq \approvaldash\calt.$
\end{enumerate}
\end{corollary}

We can extend the five-type collapse in Part~1 of 
Corollary~\ref{c:approval-axiomatic-results} to a six-type collapse.

\begin{theorem}\label{t:approval-extend-to-6-collapse}
    $\approvaldcrpctpuw = \approvaldcpctpuw$.
\end{theorem}
\begin{proof}
    We will show that both sets coincide with the set $B = \{(C, V, p) \condition p \in C$ and
$p$ is not a unique winner of approval election $(C, V)\}$.

    It follows from the fact that approval is immune with respect to $\dcrpctpuw$ and 
    $\dcpctpuw$~\cite{hem-hem-rot:j:destructive-control} that
    $\approvaldcrpctpuw$ and $\approvaldcpctpuw$
     are both subsets of $B$. Fix $(C, V, p) \in B$. 
    Then the partition $(\emptyset, C)$ witnesses that $(C, V, p) \in \approvaldcpctpuw$ as the final
    stage will be approval election $(C, V)$. Similarly, the partition $(\emptyset, C)$ also witnesses that
    $(C, V, p) \in \approvaldcrpctpuw$. This follows from the fact that $p$ is not a unique winner of the
    approval election $(C, V)$ and so there must exist a different candidate $d$ who is a winner of election $(C, V)$ and so whose score is at least as
    large as $p$'s score. No one will proceed from 
    subelection $(\emptyset, V)$ and $d$ will proceed from subelection $(C, V)$ 
    (along with those candidates, if any, who tie with $d$) and be a winner of the final round. Thus even if $p$ proceeds to the final round, it will not be a unique winner.~\end{proof}

Additionally, we prove the following immunity result about approval voting 
in the nonunique-winner model; the analogous result for the unique-winner model was obtained
by~\citeA{hem-hem-rot:j:destructive-control}.

\begin{theorem}\label{t:approval-immune-to-candidate-control}
    Approval is immune to constructive control by partition of candidates and run-off
    partition of candidates under the \tp\ tie-handling rule in the nonunique-winner model.
\end{theorem}
\begin{proof}
    We first show that a winner (possibly tied) of the two-stage approval election induced by 
    partition of candidates under the \tp\ tie-handling rule must be a winner without any control 
    action (i.e., in the input's election).
    Fix an election $(C, V)$ and let $p \in C$ be a candidate such that 
    $(C, V, p) \in \approvalccpctpnuw$. Thus it holds that $p$'s score is at least as high as the score
    of the other candidates present in the final round. Since no candidate is eliminated by the tie-handling rule, each candidate that is eliminated in the first round has score strictly less
    than some candidate that is present in the final round. It follows that $p$'s score is at least as
    high as the score of every other candidate in $C$. Thus $p$ is a winner of the approval election $(C, V)$.
    
    Essentially the same argument line can be used to show that for each $(C, V, p) \in 
    \approvalccrpctpnuw$, 
    it holds that $p$ is a winner of approval election $(C, V)$.~\end{proof}

\begin{theorem}\label{t:av-cc-c-tp-nuw}
    $\approvalccpctpnuw = \approvalccrpctpnuw$.
\end{theorem}
\begin{proof}
	We will show that
	$\approvalccpctpnuw$ and $\approvalccrpctpnuw$ are 
	comprised of exactly those inputs
	where the focus candidate is already a nonunique winner, 
	following the approach in the proofs of 
	Theorems~\ref{t:unique-alpha-collapse} and~\ref{t:unique-alpha-cc-44}.

Consider the following two disjoint sets: $A = \{(C, V, p) \condition p \in C$ and
$p$ is a winner 
of approval election $(C, V)\}$ and 
$B = \{(C, V, p) \condition p \in C$ and
$p$ is not a winner of approval election $(C, V)\}$. 
It's clear that $\approvalccpctpnuw \subseteq (A \cup B)$ and $\approvalccrpctpnuw \subseteq (A \cup B)$.
By Theorem~\ref{t:approval-immune-to-candidate-control}, it holds that
approval is immune to constructive control by partition of candidates and run-off
partition of candidates under the \tp\ tie-handling rule in the nonunique-winner model.
Thus $\approvalccpctpnuw \cap B = \emptyset$ and  $\approvalccrpctpnuw \cap B = \emptyset$, and
it holds that $\approvalccpctpnuw \subseteq A$ and $\approvalccrpctpnuw \subseteq A$.
Fix $(C, V, p) \in A$. Then the partition $(\emptyset, C)$ witnesses that both $(C, V, p) \in \approvalccpctpnuw$ and $(C, V, p) \in \approvalccrpctpnuw$ as no candidate will have score higher than $p$'s score (because $p$ is a winner of approval election $(C, V)$), and since the tie-handling rule does not eliminate candidates, $p$ will always proceed from the subelection it participates in to the final round. Thus no candidate can defeat $p$ in the final round, making $p$ a winner of the two-stage election.~\end{proof}

\begin{corollary}\label{c:av-cc-nuw-containments} %
    For each $\calt \in$ $\{\ccpctenuw$, $\ccrpctenuw$, $\ccpvtpnuw\}$,
    it holds that $\approvalccpctpnuw \subsetneq \approvaldash\calt.$
\end{corollary}
\begin{proof}
	Fix $\calt \in$ $\{\ccpctenuw$, $\ccrpctenuw$, $\ccpvtpnuw\}$
	and fix $(C, V, p) \in \approvalccpctpnuw$.
	By the proof of Theorem~\ref{t:av-cc-c-tp-nuw}, we know that 
	$\approvalccpctpnuw$ is
	comprised of only those inputs where $p$ is already a nonunique winner of the approval
	election $(C, V)$.
	In the first case, where $\calt$ is about partitioning candidates, then the partition $(\{p\}, 
	C-\{p\})$ is a witness that $(C, V, p) \in \approvaldash\calt$ as $p$ will uniquely win the 
	subelection it participates in and proceed to the final round. Regardless of which candidates 
	proceed to the final round from the other subelection, they will not have score greater than 
	$p$'s (or they would have defeated $p$ in the $(C,V)$ election) and so $p$ is a winner of the 
	two-stage election.
	In the second case, where $\calt$ is about partitioning votes (i.e., $\calt=\ccpvtpnuw$),
	then the partition $(\emptyset, V)$ suffices as everyone proceeds to the final round (because
	all the candidates tie in the subelection $(C, \emptyset)$ and the tie-handling rule does not 
	eliminate candidates) and since $p$ is already a winner of the approval election $(C, V)$, they 
	are a winner 
	of the final round. The corresponding separation witness can be found in 
	Table~\ref{table:approval-results}.~\end{proof}

Finally, we provide direct arguments that yield additional collapses and containments.
\begin{theorem}\label{t:av-dc-pv-te} %
	$\approvaldcpvteuw = \approvaldcpvtenuw$.
\end{theorem}
\begin{proof}
	The $\supseteq$ relationship is immediate.

 	$\subseteq$: Let $(C, V, p) \in
	\approvaldcpvteuw$.  Note that this 
  implies  $\|C\| \geq 2$, since in one-candidate approval
  elections the one candidate always wins.
 Let $(V_1, V_2)$ be a vote partition that witnesses $(C, V, p) \in
	\approvaldcpvteuw$.  
	As the tie-handling rule is the \te\ rule,
	it must hold (since if $p$ uniquely won both subelections it necessarily would uniquely win
	the second-round election)
	that either $p$ is eliminated in both subelections
	(either by tieing or by losing)
	or
	$p$ uniquely wins one subelection, some other candidate $d$ uniquely wins
	the other subelection, and $p$ does not uniquely win the final round
	(thus $d$'s score must be at least as large as $p$'s score when using vote set $V$).
	In the first case, the partition $(V_1, V_2)$ witnesses
	$(C, V, p) \in \approvaldcpvtenuw$ since $p$ will not proceed to the final round.
	In the second case, using the partition $(V, \emptyset)$ suffices for the following
	reasons. In the subelection $(C, V)$, since $d$'s score is at least as large as that of $p$,
	either $d$ uniquely wins the subelection or both $p$ and $d$ tie and are eliminated by the tie-handling
	rule. In the subelection
	$(C, \emptyset)$ every candidate ties and so all candidates are eliminated by the tie-handling 
	rule (since $\|C\| \geq 2$).
	Thus $p$ is eliminated in both subelections, does not proceed to the final round,
	and so is not a final-round winner.~\end{proof}

\begin{theorem}\label{t:av-cc-c-te-nuw} %
	$\approvalccpctenuw = \approvalccrpctenuw$.
\end{theorem}
\begin{proof}

    We structure this proof so as to yield not just this theorem but also the related result that we state as Corollary~\ref{c:av-cc-c-te-uw}.

    $\subseteq$: Let $(C, V, p) \in \approvalccpctenuw$ and let $(C_1, C_2)$ be a candidate 
    partition that witnesses this membership. 
    Consider the case where $p$ uniquely wins the final round. If $p \in C_1$, then $p$
    uniquely wins $(C_1, V)$ and in the final round defeats all candidates in $C_2$. If $p \in 
    C_2$, then in the final round
    $p$ defeats the candidate (if any) that survives the \te\ tie-handling rule regarding the 
   	subelection $(C_1, V)$ as well as all the candidates in $C_2 - \{p\}$.
    Regardless of which case holds, the partition $(C_1, C_2)$ will witness
    $(C, V, p) \in \approvalccrpctenuw$ since $p$ will 
    uniquely win its subelection and then will defeat any candidate that moves forward from the 
    other subelection.
    If $p$ does not uniquely win the final round, then there is at least one other candidate that 
    ties with $p$ in the final round.
    If $p \in C_1$, then $p$ must uniquely win in $(C_1, V)$ and as no candidate 
    in $C_2$ can have a score greater than $p$'s (since $(C_1, C_2)$ witnesses 
    $(C, V, p) \in \approvalccpctenuw$), the partition $(C_1, C_2)$ suffices
    to witness $(C, V, p) \in \approvalccrpctenuw$\@.
    If $p \in C_2$, then under partition $(C_1, C_2)$ $p$ could first-round tie with a candidate 
    and be eliminated (under run-off
    partition of candidates due to the \te\ rule).
    However, let $T$ denote the (possibly empty) set of candidates (other than $p$) that tie with 
    $p$ in $(C_2, V)$.
    Then the partition $(C_1 \cup T, C_2 - T)$ witnesses $(C, V, p) \in \approvalccrpctenuw$, 
    since $p$ will uniquely win $(C_2 - T, V)$ and will either tie
    or defeat the winner (if any) of $(C_1 \cup T, V)$.
	
	$\supseteq$: Let $(C, V, p) \in \approvalccrpctenuw$ and let $(C_1, C_2)$ be a candidate
	partition that witnesses this membership.
	Without loss of generality, assume that $p \in C_1$.
	Thus it holds that $p$ uniquely wins $(C_1, V)$.
	If $p$ uniquely wins the final round, then $p$ also defeats the candidate (if any) that moves
	forward from $(C_2, V)$. Thus the partition $(C_2, C_1)$ will also witness $(C, V, p) \in 
	\approvalccpctenuw$ (since $p$ has strictly more approvals than any candidate other than 
	itself).
	If $p$ does not uniquely win the final round, then there is another candidate $d$, who
	is the unique winner of $(C_2, V)$ and ties with $p$ in the final round. 
	Again the partition	$(C_2, C_1)$ suffices to witness $(C, V, p) \in \approvalccpctenuw$ since 
	$d$ proceeds to the final round and both $p$ and $d$ win there due to their numbers of 
	approvals.~\end{proof}

\begin{corollarytotheproof}\label{c:av-cc-c-te-uw} %
$\approvalccpcteuw = \approvalccrpcteuw$.
\end{corollarytotheproof}
\begin{proof}
	This is an immediate corollary to the above proof of Theorem~\ref{t:av-cc-c-te-nuw}, as the 
	proof was intentionally structured to ensure that if the witness of one type made $p$ a unique 
	winner of the final round, then the constructed-above (sometimes different) partition for the 
	other type also made $p$ a unique winner of the final round under that other type of 
	control.~\end{proof}

We can additionally show the following 
three
containments.
Theorem~\ref{t:av-dcrpctenuw-in-dcpvtpuw}
and Corollary~\ref{c:av-dcrpctenuw-in-dcpvtenuw} are surprising, since they are about partitioning
different components of elections.

\begin{theorem}\label{t:av-dcpvtpuw-in-dcpvtenuw}
    $\approvaldcpvtpuw \subsetneq \approvaldcpvtenuw$.
\end{theorem}
\begin{proof}
The proof is similar in flavor to that
of Theorem~\ref{t:av-dc-pv-te}.
Let $(C, V, p) \in \approvaldcpvtpuw$ and let $(V_1, V_2)$ be a vote partition that witnesses this
membership.
There are two cases to consider. Either 
$p$ uniquely wins in exactly one subelection (it certainly cannot uniquely win in both, else $p$ would be the unique final-round winner), or 
$p$ is not a unique winner in either subelection.
In the latter case, the partition $(V_1, V_2)$ witnesses
that $(C, V, p) \in \approvaldcpvtenuw$ as $p$ never
survives the \te\ tie-handling rule (because it is not a unique winner in either subelection). In the former
case, the partition $(V, \emptyset)$ witnesses
that $(C, V, p) \in \approvaldcpvtenuw$. 
This is so because, in that case, there must exist a candidate $d \neq p$ 
that 
in the $\approvaldcpvtpuw$ setting under partition $(V_1,V_2)$
proceeds
to the final stage and 
in the second round has a score at least as high as that 
of $p$. 
So in the $\approvaldcpvtenuw$ setting's first round
under the partition $(V,\emptyset)$, 
in the subelection $(C, V)$ candidate $p$ can at best tie $d$,
and thus 
certainly cannot move forward under the \te\ tie-handling rule.
Additionally, no one moves forward from subelection
$(C, \emptyset)$ 
(everyone ties and, since in the current case we 
know there are at least two candidates, 
everyone is eliminated from that subelection).
This shows the containment. The separation witness 
for the strict containment
can be found in Table~\ref{table:approval-results}.~\end{proof}

\begin{theorem}\label{t:av-dcrpctenuw-in-dcpvtpuw}
    $\approvaldcrpctenuw \subsetneq \approvaldcpvtpuw$.
\end{theorem}
\begin{proof}
    Let $(C, V, p) \in \approvaldcrpctenuw$ and let $(C_1, C_2)$ be a candidate partition that 
    witnesses this membership. Then, with respect to that partition, $p$ is eliminated either in 
    its subelection (either by tieing or by losing)
    or in the final round. In both cases, this must happen because there is another candidate $d$
    such that $p$'s score is at most $d$'s score (using vote set $V$). 
    By using the partition $(\emptyset, V)$, we 
    know that every candidate will proceed to the final round (everyone ties in $(C, \emptyset)$ and proceeds to the final round)
    and in that final round, either
    $d$ will tie $p$ or $d$ will defeat $p$. In either case, $p$ is not a unique winner.
    This shows the containment. The separation witness for the strict containment can be
found in Table~\ref{table:approval-results}.~\end{proof}

As an immediate corollary to Theorems~\ref{t:av-dcpvtpuw-in-dcpvtenuw} 
and~\ref{t:av-dcrpctenuw-in-dcpvtpuw}, we have the following result.
\begin{corollary}\label{c:av-dcrpctenuw-in-dcpvtenuw}
    $\approvaldcrpctenuw \subsetneq \approvaldcpvtenuw$.
\end{corollary}

\section{Related Work}

Electoral control attacks were introduced and studied by
Bartholdi, Tovey, and Trick in a tremendously influential
1992 paper~\cite{bar-tov-tri:j:control}.
They defined a set of attacks based on adding/deleting/partitioning
candidates and voters, and those attacks have been studied ever since.

\citeA{bar-tov-tri:j:control} studied only the goal of making a particular
focus candidate be an untied winner of the election (i.e., constructive control in the unique-winner model).
\citeA{hem-hem-rot:j:destructive-control}
introduced ``destructive control'' versions of each constructive control
type: The goal is to prevent (via the given control action) the focus
candidate from becoming a unique winner. 
Later papers
often additionally, or only, used the nonunique-winner model, in which
the goal is to make the focus candidate be (or, for the destructive case, not 
be) a winner.
As far as we know, the first example of this was a study 
of Llull and Copeland elections by 
\shortciteA{fal-hem-hem-rot:cOutByJour:llull,fal-hem-hem-rot:j:llull}
that studied both
winner 
models, 
and 
\citeA{hem-lav-men:j:schulze-and-ranked-pairs} 
(see also~\citeR[Footnote~5]{hem-hem-men:j:search-versus-decision})
argue that
the nonunique-winner model is a better model to study than the unique-winner
model.
\citeA{hem-hem-rot:j:destructive-control}
addressed the ways of handling ties in the first-round elections of the (two-round)
``partition'' control types of 
\citeA{bar-tov-tri:j:control}, and framed the two now-standard
approaches called ``ties promote'' and ``ties eliminate''; 
they 
did this 
because although tie-handling had previously been suggested as being unimportant,
their paper establishes that, for example, for plurality the choice between 
those two rules spells the 
difference between being NP-complete and belonging to P\@.  The ``adding candidates''
control attack of 
\citeA{bar-tov-tri:j:control} was anomalously defined, and
\citeA{hem-hem-rot:j:hybrid} kept
the original notion but renamed it ``unlimited adding of candidates,''
and introduced, under the thus-available name ``adding candidates,'' the
version that is analogous to the other 
\citeA{bar-tov-tri:j:control} add/delete
types, and the subsequent papers have followed that notational shift.

Altogether, the 
\citeA{bar-tov-tri:j:control} control attack set,
under the above clarifications and enrichments, yields a set of 
eleven constructive control attacks and 
eleven
destructive control attacks. As mentioned earlier, this in some sense forms a 
``standard'' benchmark set of attacks
(though some papers use subsets of that collection, and other papers
have taken control in additional directions, e.g., resolute control~\citeR{wan-yan:c:resolute-control-aamas-2017-anyone-but-them-yang-wang,gup-roy-sau-zeh:j:resolute-control}).
For example, the excellent survey 
chapter
on
control and bribery by 
\citeA{fal-rot:b:handbook-comsoc-control-and-bribery} uses precisely
those 22 control attacks,
as does the recent paper on search versus decision of 
\citeA{hem-hem-men:j:search-versus-decision}.
Since the field has not yet resolved whether nonunique-winner
or
unique-winner 
is the right standard---indeed, the two just-cited
sources make different choices---and because discovering cases when a 
control type in one of those models turns out to be identical to a
control type in the other is itself interesting, in this paper we covered
both models, and thus
$2 \times (11 + 11) = 44$ control types.

Of the $\binom{44}{2} = 946$ pairs (322 of them compatible) of those control types, seven were
proven to collapse in 
\citeA{hem-hem-men:j:search-versus-decision}. That is the paper
that most strongly inspired the present paper.  It also 
separates one pair: \dcrpctpuw\ and \dcpctpuw. However, it 
does so 
using an election system that is
not \mbox{(candidate-)neutral}. 
In our 
present paper,  regarding 
that separation and all separations, 
we obtain 
them directly in 
systems that are \mbox{(candidate-)neutral}
or by inheritance
from systems that are \mbox{(candidate-)neutral}.

The additional, different control types known as resolute control~\cite{wan-yan:c:resolute-control-aamas-2017-anyone-but-them-yang-wang,gup-roy-sau-zeh:j:resolute-control}
are quite interesting.  Resolute control asks whether 
there is some action (from a certain 
range of actions) that keeps every one of a certain
collection of candidates from being a winner.  This might
seem to be the same as our function model for the case of 
(nonunique-winner) destructive control, but it is not.  
In our function model for nonunique-winner destructive 
control, we are 
speaking of the collection of candidates who can individually
be prevented from winning by some control action.  But even 
if two or more candidates belong 
to our function's output, they could be
put into that output by different control actions, and 
there might be no single action that blocks both 
simultaneously.  In brief, resolute control is focused 
on blocking whole groups, and our function model in contrast
is a refinement of set separations and focuses on
individual candidates to let us identify new
containment patterns between control types.

Collapsing or separating control types is not \emph{directly} about
complexity. However, doing so is highly relevant to complexity, as the
types were defined as natural benchmarks whose complexity could be studied.
In fact, there has been something of a race to find natural systems in which
very many control attacks are $\np$-hard to carry 
out.\footnote{%
Regarding the ``race'' to 
find systems that are resistant to a large number of control types, one 
might ask whether it still is fair to count it as multiple strengths 
if the elements of collapsing pairs (or larger equivalence classes)
are resistant
(and so NP-hard), or to count it as multiple weaknesses if 
the elements of collapsing pairs (or larger equivalence classes)
are vulnerable (and thus in P).  
In some sense, one could argue 
that this is unfair, as it is putting extra weight on the ``same''
type.  

However, one could also argue that having a collapsing 
control type foursome count as four is fair and natural, since 
the actual mechanisms of the four are different, and so one 
either is fighting off attacks through NP-completeness, or is vulnerable to
multiple attack lines.  

In any case, if one focuses solely on  
number of resistances, one is implicitly saying that all control 
types are equally important, and we do not think that saying that 
is correct.  Rather, what is most important is that the field knows, for the full palette of control types, which ones are resistant and which ones are vulnerable with respect to a given election system.  That will allow the people choosing which election system to use in a given setting to choose an election system that is resistant to as many as possible of the attacks that they expect are the most likely within the setting. And for that, there is no doubt that avoiding duplicate work through exploiting collapses is helpful.

Finally, we mention that merging collapsed types on each system's 
``score card'' would be problematic, since different elections have 
different collapses, and so different election
systems' score cards would not even have the same 
number of entries.}
Among the many
systems that have done well or very well in that race are, along with some
of the key papers that analyzed the complexity of each, 
Schulze elections~\cite{par-xia:c:ranked-pairs,men-sin:c:schulze,hem-lav-men:j:schulze-and-ranked-pairs},
ranked-pair elections~\cite{par-xia:c:ranked-pairs,hem-lav-men:j:schulze-and-ranked-pairs},
SP-AV elections~\cite{erd-now-rot:j:sp-av}, normalized range 
voting~\cite{men:j:range-voting}, fallback 
elections (\citeR{erd-rot:c:fallback,erd-pir-rot:c:open-problems},
see also~\citeR{erd-fel-rot-sch:j:control-in-bucklin-and-fallback-voting}),
and Bucklin elections (\citeR{erd-pir-rot:c:open-problems},
see also~\citeR{erd-fel-rot-sch:j:control-in-bucklin-and-fallback-voting}).
\citeA[Table~7.3]{fal-rot:b:handbook-comsoc-control-and-bribery}
provide a lovely table, for the 22 unique-winner control cases, of what
is known for twelve voting systems. Regarding the three important systems
we spotlight in the present paper, plurality's control was explored by
\citeA{bar-tov-tri:j:control} and
\citeA{hem-hem-rot:j:destructive-control},
veto
has been studied by 
\citeA{lin:thesis:elections} 
and 
\citeA{mau-rot:j:control-veto-plurality}
(see also~\citeR[Table~1]{erd-reg-yan:c:puzzle}), and
approval has been studied by~\citeA{hem-hem-rot:j:destructive-control}
(see also~\citeR{bau-erd-hem-hem-rot:b:approval}).

\section{%
Comments on Checking and Reproducibility}

As mentioned earlier, a number of our election examples showing 
separations were found with the aid of computer searches.
Other examples were human generated.

In addition to the verification code within 
the computer-search programs, we coded---with a different
person doing the coding---stand-alone verification
routines, as a double-check of the correctness of all separation examples used in
this paper.
Whether computer-generated or human-generated, every example
was checked via those separate verification programs.
Also, 
every computer-generated example was %
verified by a human.

To support reproducibility, and to aid researchers who might wish to carry our study to other cases, we have made the computer-search programs, and their inputs and outputs, publicly available, in effect as an online appendix of this paper, in an online repository. That repository is available at 
\mbox{\url{https://github.com/MikeChav/SCT_Code}}.

Our search programs use randomization, and the repository captures and documents the randomization used in each run generating our examples.  Thus skeptical researchers could, if they wanted, simulate our code, using the same randomization, to assure themselves that our codes indeed produced the examples we claim they did.  The repository also includes some information about what systems the programs were run on and how long certain runs took.
(In fact, each program was run once, and the run itself randomly tried many examples until a counterexample was found; even the longest-running program ran for at most 21 CPU seconds.)

The repository additionally includes the separate verification programs mentioned above.

\section{Conclusions and Open Problems}

Table~\ref{t:summary} summarizes our results.
We established that in the general (universally quantified) case there are no 
collapsing pairs (by which we always
mean among the standard 44 control types) other than the seven 
collapsing pairs identified by 
\citeA{hem-hem-men:j:search-versus-decision}, and that plurality has no collapsing 
pairs beyond those seven. 
For veto and approval voting we discovered additional collapsing pairs beyond those
inherited seven, but we also established that veto and approval voting, after our 
work, have no remaining 
undiscovered collapsing pairs.

Our work helps clarify the landscape of which control pairs do or do not collapse, both 
in the general (universally quantified) case, and for plurality, veto, and approval voting.

We also refined all our separations, and in doing so uncovered containment 
relationships---including many that do not follow from the relationship 
between the nonunique-winner model and the unique-winner model.
\begin{table}[!tb]
\caption{\protect\label{t:summary} Summary of separations and collapses.  
 {\color{blue} Blue} indicates results due to
  or inherited from~\cite{hem-hem-men:j:search-versus-decision}. 
   {\color{purple} Red}
  indicates results due to the present paper.
  The general-case line shows when collapses occur for
  all election systems over 
  linear orders.%
}
\small
\centering
\begin{tabular}{l| c c  c | c c c }
 & \multicolumn{3}{c|}{~~~Set Classification} & \multicolumn{3}{c}{Subclassification of Separations} \\ 
Election System          & Separations & Collapses & Open & ``$\subsetneq$''/``$\supsetneq$''  &  Incomparable & Open\\ \hline
   General Case    &
{\color{blue} 1} +               {\color{purple}  314} = 315${}^\dagger$
 & {\color{blue} 7} 
& 0  &  {\color{purple}  38} &  {\color{purple}  277} &  0\\ 
   Plurality  &                      {\color{purple}  315}
  & {\color{blue} 7}  & 0 &{\color{purple}  38} &  {\color{purple}  277} &  0  \\ 
   Veto        &
                     {\color{purple}  314}
 & {\color{blue} 7} + {\color{purple} 1} = 8 & 0 & {\color{purple}  58} &  {\color{purple}  256} & 0  \\
   Approval Voting &
{\color{purple}  301} & {\color{blue} 7} + {\color{purple} 14} = 21
 & 0 & {\color{purple} 88} & {\color{purple} 213}  &0 %
\end{tabular}\\
\raggedright
${}^\dagger$Or
   {\color{blue} 0} +               {\color{purple}  315  }
  for the pure social choice approach to candidate 
  names (see Footnote~\ref{f:names} and
  the discussion in the Related Work section).
\end{table}

A number of interesting open directions are suggested by our work. One is, for important 
concrete election systems other than plurality, veto, and approval voting, to completely 
classify the collapses and separations that hold for those specific systems. Another 
direction---building on 
the results of Section~\ref{sub:av}---is 
to find sufficient conditions (or, better still, necessary-and-sufficient conditions) for many 
control-pair collapses in terms of 
axiomatic properties of election systems.
Though our goal for separations was to subclassify each
separation into exactly one of the three cases 
``$\subsetneq$'', ``$\supsetneq$'', or incomparability, in our 
tables we have also 
noted those cases where our constructions establish 
strong incomparability; one could for the cases where 
we list incomparability try to establish strong 
incomparability. %
An additional open direction is to see whether already-studied control types beyond the 44 investigated 
here  collapse with each other or with some of the 44, either in general or for important 
concrete election systems.

Finally, control types are defined as sets. When a pair of control types collapses,
those sets are equal and thus certainly are of the same complexity. However, it would
be interesting to see whether for collapsing control types, their complexities as
\emph{search} problems are or are not polynomially related.
That issue, inspired by the work of 
\citeA{hem-hem-men:j:search-versus-decision}
and an earlier version of the present 
paper, has recently been studied 
by 
\citeA{car-cha-hem-nar-tal-wel:c-aamas:search-vs-search}.

\paragraph{Acknowledgments}
Some of these results appeared in the \emph{Proceedings of the 22nd International Conference on Autonomous Agents and Multiagent Systems}~\cite{car-cha-hem-nar-tal-wel:c:sct}. We thank
anonymous 
referees 
and 
Ulle Endriss 
for their 
valuable comments, suggestions, and guidance.

%

%
%
%
\bibliographystyle{theapa}

\appendix

\section{Compatible Control Types}
Table~\ref{table:compatibility} shows how the 44 standard control types 
partition into five compatibility equivalence classes.  Every other (distinct)
pair among the 44 
is not a compatible pair.

\medskip
{

\smalltblfont
\begin{longtable}{c|c|c|P{0.5in}|P{0.5in}|P{0.4in}|P{0.5in}}%
\caption{The 44 types of control and a description of which components are part of the input for each one. The input type thus partitions the control types into five equivalence classes as to compatibility of inputs.}\\
\label{table:compatibility}
\centering
Control Type&Candidates&{Votes}&Focus Candidate&Spoiler Candidates&Spoiler Votes&Limit (Natural Number)\\
\endfirsthead
Control Type&Candidates&{Votes}&Focus Candidate&Spoiler Candidates&Spoiler Votes&Limit (Natural Number)\\
\endhead
\hline
CC-PV-TE-UW, CC-PV-TE-NUW,&\multirow[c]{12}{*}{Yes}&\multirow[c]{12}{*}{Yes}&\multirow[c]{12}{*}{Yes}&\multirow[c]{12}{*}{No}&\multirow[c]{12}{*}{No}&\multirow[c]{12}{*}{No}\\
CC-PV-TP-UW, CC-PV-TP-NUW,&&&&&&\\
CC-PC-TE-UW, CC-PC-TE-NUW,&&&&&&\\
CC-PC-TP-UW, CC-PC-TP-NUW,&&&&&&\\
CC-RPC-TE-UW, CC-RPC-TE-NUW,&&&&&&\\
CC-RPC-TP-UW, CC-RPC-TP-NUW,&&&&&&\\
DC-PV-TE-UW, DC-PV-TE-NUW,&&&&&&\\
DC-PV-TP-UW, DC-PV-TP-NUW,&&&&&&\\
DC-PC-TE-UW, DC-PC-TE-NUW,&&&&&&\\
DC-PC-TP-UW, DC-PC-TP-NUW,&&&&&&\\
DC-RPC-TE-UW, DC-RPC-TE-NUW,&&&&&&\\
DC-RPC-TP-UW, DC-RPC-TP-NUW&&&&&&\\
\hline
CC-AC-UW, CC-AC-NUW,&\multirow[c]{2}{*}{Yes}&\multirow[c]{2}{*}{Yes}&\multirow[c]{2}{*}{Yes}&\multirow[c]{2}{*}{Yes}&\multirow[c]{2}{*}{No}&\multirow[c]{2}{*}{Yes}\\
DC-AC-UW, DC-AC-NUW&&&&&&\\
\hline
CC-DC-UW, CC-DC-NUW,&\multirow[c]{4}{*}{Yes}&\multirow[c]{4}{*}{Yes}&\multirow[c]{4}{*}{Yes}&\multirow[c]{4}{*}{No}&\multirow[c]{4}{*}{No}&\multirow[c]{4}{*}{Yes}\\
CC-DV-UW, CC-DV-NUW,&&&&&&\\
DC-DC-UW, DC-DC-NUW,&&&&&&\\
DC-DV-UW, DC-DV-NUW&&&&&&\\
\hline
CC-AV-UW, CC-AV-NUW,&\multirow[c]{2}{*}{Yes}&\multirow[c]{2}{*}{Yes}&\multirow[c]{2}{*}{Yes}&\multirow[c]{2}{*}{No}&\multirow[c]{2}{*}{Yes}&\multirow[c]{2}{*}{Yes}\\
DC-AV-UW, DC-AV-NUW&&&&&&\\
\hline
CC-UAC-UW, CC-UAC-NUW,&\multirow[c]{2}{*}{Yes}&\multirow[c]{2}{*}{Yes}&\multirow[c]{2}{*}{Yes}&\multirow[c]{2}{*}{Yes}&\multirow[c]{2}{*}{No}&\multirow[c]{2}{*}{No}\\
DC-UAC-UW, DC-UAC-NUW&&&&&&\\
\\
\end{longtable}

}

\section{Tables}

Tables~\ref{table:plurality-tools},~\ref{table:veto-tools},~and~\ref{table:approval-tools}
list each of the elections (identified by a candidate set $C$ and a vote set $V$) used 
to show separations and incomparability. Each line of these tables
also has some columns that may have no entries (denoted by ``-'').
These columns are $S$ (to denote additional candidates when the control
type is about adding candidates), $U$ (to denote additional votes when the control type is about adding votes), and $k$ (to denote the limit imposed on a particular control, e.g., 
limited adding of candidates, or deleting candidates/voters).

We assign an ID to each election and use those IDs from 
Tables~\ref{table:plurality-tools},~\ref{table:veto-tools},~and~\ref{table:approval-tools} in Tables~\ref{table:plurality-results},~\ref{table:veto-results},~and~\ref{table:approval-results}
to identify which tool is a separation witness.
The obvious containments (that are about UW and NUW variants of the same control type,
see 
location $(**)$ in Section~\ref{sec:introduction})
are implicit in this table and thus we do not explicitly mark when they are used.

Additionally, we color code
Tables~\ref{table:plurality-results}, \ref{table:veto-results}, and~\ref{table:approval-results} 
in a specific manner: The entries that have the 
same (nonwhite) background color form an equivalence class.
Each entry that is not colored (i.e., that has a white background) is an equivalence class of size one, and we will not color such entries. Table~\ref{table:equivalence-classes} provides a summary of the equivalence classes and their corresponding colors. 

Each boldfaced entry indicates the canonical element of its equivalence class. 
Those are also
indicated in Tables~\ref{table:plurality-results},~\ref{table:veto-results},~and~\ref{table:approval-results}.
Having canonical elements helped make the proof process
more economical. 
If we separate $\calt$ from 
$\calt'$, we implicitly have separated $\calt$ from
all members of the equivalence class of $\calt'$. In
the proof process we focused on separations regarding
only two canonical elements (where we view each singleton
type as a stand-alone canonical element of its 
size-one equivalence class, though we don't boldface 
those).

\newcommand{\pluralityclassone}{\cellcolor{gray!50}}
\newcommand{\pluralityclasstwo}{\cellcolor{blue!50}}

\newcommand{\vetoclassone}{\cellcolor{green!25}}
\newcommand{\vetoclasstwo}{\cellcolor{blue!25}}
\newcommand{\vetoclassthree}{\cellcolor{cyan}}

\newcommand{\approvalclassone}{\cellcolor{olive!50}}
\newcommand{\approvalclasstwo}{\cellcolor{teal!50}}
\newcommand{\approvalclassthree}{\cellcolor{red!50}}
\newcommand{\approvalclassfour}{\cellcolor{purple!50}}
\newcommand{\approvalclassfive}{\cellcolor{orange!90}}
\newcommand{\approvalclasssix}{\cellcolor{pink}}
\newcommand{\approvalclassseven}{\cellcolor{brown!75}}

\medskip

\begin{longtable}{p{5in}|c}
\caption{Equivalence classes and their respective colors. Each boldfaced entry indicates the canonical element of its equivalence class.}\\
\label{table:equivalence-classes}
\centering
Class & Color \\\hline
\endfirsthead
\centering Class & Color \\\hline
\endhead
\raggedright {\bf Plurality-DC-RPC-TE-NUW}, 
$\pluralitydcrpcteuw$, $\pluralitydcpcteuw$,  $\pluralitydcpctenuw$ & \pluralityclassone\\ \hline
\raggedright {\bf Plurality-DC-RPC-TP-NUW}, $\pluralitydcpctpnuw$ & \pluralityclasstwo\\ \hline
\raggedright {\bf Veto-DC-RPC-TE-NUW}, $\vetodcrpcteuw$,  
$\vetodcpcteuw$, $\vetodcpctenuw$ & \vetoclassone\\ \hline
\raggedright {\bf Veto-DC-RPC-TP-NUW}, $\vetodcpctpnuw$ & \vetoclasstwo\\ \hline
{\raggedright \bf Veto-DC-PV-TE-NUW}, $\vetodcpvteuw$ & \vetoclassthree\\ \hline
\raggedright {\bf Approval-DC-PV-TE-NUW}, $\approvaldcpvteuw$ & \approvalclassone\\ \hline
\raggedright {\bf Approval-DC-RPC-TE-NUW}, $\approvaldcrpcteuw$,
$\approvaldcpcteuw$,
$\approvaldcpctenuw$,
$\approvaldcrpctpuw$, $\approvaldcpctpuw$& \approvalclasstwo\\ \hline
\raggedright {\bf Approval-DC-RPC-TP-NUW}, $\approvaldcpctpnuw$ & \approvalclassthree\\ \hline
\raggedright {\bf Approval-CC-RPC-TP-UW}, $\approvalccpctpuw$ & \approvalclassfour\\ \hline
\raggedright {\bf Approval-CC-RPC-TP-NUW}, $\approvalccpctpnuw$ & \approvalclassfive\\ \hline
\raggedright {\bf Approval-CC-RPC-TE-UW}, $\approvalccpcteuw$ & \approvalclasssix\\ \hline
\raggedright {\bf Approval-CC-RPC-TE-NUW}, $\approvalccpctenuw$ & \approvalclassseven\\ 
\end{longtable}
\medskip
{
\newcommand{\plurVwidth}{4.5in}

\smalltblfont
\setlength{\tabcolsep}{2pt}
\begin{longtable}[c]{c|P{0.5in}|c|p{\plurVwidth}|c|c}
\caption{List of separation witnesses in plurality. We note the computer-generated entries with a ``$\dagger$'' superscript.}\\
\label{table:plurality-tools}
ID & $C$ & $S$ & \parbox[c]{\plurVwidth}{\centering $V$} & $U$ & $k$\\\hline
\endfirsthead
ID & $C$ & $S$ & \parbox[c]{\plurVwidth}{\centering $V$} & $U$ & $k$\\\hline
\endhead
Plur.1 & $\{a, b, c\}$ & - & $\{a>b>c, b>a>c, c>a>b\}$ & - & - \\
Plur.2 & $\{a, b\}$ & - & $\{a>b, b>a\}$ & - & - \\
Plur.3 & $\{a, b\}$ & - & $\{a>b\}$ & - & - \\
Plur.4${}^\dagger$ & $\{a,b,c, d\}$ & - & $\{b>c>d>a, d>a>c>b, b>c>d>a, a>c>b>d, a>b>d>c, d>a>b>c, c>d>b>a, d>a>c>b, a>c>b>d, d>c>b>a, b>c>d>a, a>b>d>c, d>b>c>a, a>d>c>b, b>c>d>a, c>a>b>d, b>a>d>c, a>c>d>b\}$ & - & - \\
Plur.5 & $\{a,b,c\}$ & - & $\{a>b>c,a>b>c,a>c>b,a>c>b, b>a>c,b>a>c,b>a>c, c>a>b,c>a>b,c>a>b\}$ & - & -\\
Plur.6 & $\{a, b, c\}$ & - & $\{b>c>a, c>b>a\}$ & - & - \\
Plur.7 & $\{a, b, c, d\}$ & - & $\{a>b>c>d, c>d>a>b, d>b>a>c\}$ & - & - \\
Plur.8 & $\{a, b, c\}$ & - & $\{a>b>c, a>b>c, b>a>c, b>a>c, c>b>a\}$ & - & - \\
\end{longtable}

}

\newcommand{\vetoVwdith}{4.3in}
\medskip

{
\smalltblfont
\setlength{\tabcolsep}{2pt}
\begin{longtable}{c|P{0.5in}|c|p{\vetoVwdith}|c|c}
\caption{List of separation witnesses in veto.
We note the computer-generated entries with a ``$\dagger$'' superscript.}\\
\label{table:veto-tools}
ID & $C$ & $S$ & \parbox[c]{\vetoVwdith}{\centering $V$} & $U$ & $k$
\\\hline\endfirsthead
ID & $C$ & $S$ & \parbox[c]{\vetoVwdith}{\centering $V$} & $U$ & $k$
\\\hline\endhead
Veto.1 & $\{a, b\}$ & - & $\{a>b, b>a\}$ & - & - \\
Veto.2 & $\{a, b, c\}$ & - & $\{a>b>c, a>b>c\}$ & - & - \\
Veto.3 & $\{a, b, c\}$ & - & $\{a>b>c, c>a>b, c>b>a, c>b>a\}$ & - & - \\
Veto.4 & $\{a, b, c\}$ & - & $\{a>b>c\}$ & - & - \\
Veto.5 & $\{a, b\}$ & - & $\{a>b\}$ & - & - \\
Veto.6 & $\{a, b\}$ & - & $\{b>a\}$ & - & - \\
Veto.7 & $\{a, b, c\}$ & - & $\{a>b>c, a>c>b, b>c>a, b>c>a\}$ & - & - \\
Veto.8 & $\{a, b, c\}$ & - & $\{a>b>c, a>b>c, c>a>b, c>b>a, c>b>a\}$ & - & - \\
\end{longtable}

}

\medskip
{

\smalltblfont
\setlength{\tabcolsep}{2pt}
\newcommand{\approvalVwidth}{4.7in}
\begin{longtable}[c]{c|P{0.5in}|c|p{\approvalVwidth}|c|c}
\caption{List of separation witnesses in approval voting. (Example Appr.7 can also be
found on page 220 of the Handbook of Approval 
Voting, \citeR{bau-erd-hem-hem-rot:b:approval}.)}\\
\label{table:approval-tools}
ID & $C$ & $S$ & \parbox[c]{\approvalVwidth}{\centering $V$} & $U$ & $k$\\\hline
\endfirsthead%
ID & $C$ & $S$ & \parbox[c]{\approvalVwidth}{\centering $V$} & $U$ & $k$\\\hline
\endhead%
Appr.1 & $\{a, b\}$ & - & $\{10\}$ & - & -\\
Appr.2 & $\{a, b\}$  & - & $\{10, 01\}$ & - & -\\
Appr.3 & $\{a, b, c\}$ & - & $\{101, 110\}$ & - & -\\
Appr.4 & $\{a, b, c\}$ & - & $\{110,110,010,\allowbreak101,\allowbreak101,\allowbreak 001\}$ & - & -\\
Appr.5 & $\{a, b, c\}$ & - & $\{100, 011, 011\}$ & - & -\\
Appr.6 & $\{a, b, c\}$ & - & $\{100, 110, 011, \allowbreak011\}$ & - & -\\
Appr.7 & $\{a, b, c\}$ & - &  $\{100$, $100$, $100$, $100$, $100$, $110$, $010$, $010$, $010$, $010$, $001$,
$001$, $001$, $001$, $001$, $001$, $001\}$ & - & -\\
Appr.8 & $\{a, b, c\}$ & - & $\{100$, $100$, $100$, $100$, $010$, $010$, $010$, $010$, $010$, $001$,
$001$, $001\}$ & - & -\\
\end{longtable}%

}

\end{document}